\pdfoutput=1
\PassOptionsToPackage{unicode}{hyperref}
\PassOptionsToPackage{hyphens}{url}
\documentclass[
]{article}
\usepackage{amsmath,amssymb}
\usepackage{iftex}
\ifPDFTeX
  \usepackage[T1]{fontenc}
  \usepackage[utf8]{inputenc}
  \usepackage{textcomp} 
\else 
  \usepackage{unicode-math} 
  \defaultfontfeatures{Scale=MatchLowercase}
  \defaultfontfeatures[\rmfamily]{Ligatures=TeX,Scale=1}
\fi
\usepackage{lmodern}
\ifPDFTeX\else
\fi
\IfFileExists{upquote.sty}{\usepackage{upquote}}{}
\IfFileExists{microtype.sty}{
  \usepackage[]{microtype}
  \UseMicrotypeSet[protrusion]{basicmath} 
}{}
\makeatletter
\@ifundefined{KOMAClassName}{
  \IfFileExists{parskip.sty}{%
    \usepackage{parskip}
  }{
    \setlength{\parindent}{0pt}
    \setlength{\parskip}{6pt plus 2pt minus 1pt}}
}{
  \KOMAoptions{parskip=half}}
\makeatother
\usepackage{xcolor}
\usepackage[margin=2.5cm]{geometry}
\usepackage{graphicx}
\usepackage{amsmath,amssymb,mathtools,bm,booktabs}
\usepackage{amsthm}

\theoremstyle{remark}
\newtheorem{remark}{Remark}
\ifLuaTeX
  \usepackage{selnolig}  
\fi
\IfFileExists{bookmark.sty}{\usepackage{bookmark}}{\usepackage{hyperref}}
\IfFileExists{xurl.sty}{\usepackage{xurl}}{} 
\hypersetup{
  pdftitle={Checking the Cumulative Exposure Model: Cumulative Chi-Squared Goodness-of-Fit Tests for Interval-Monitored Step-Stress Life Tests},
  pdfauthor={Nirian Martín},
  hidelinks,
  pdfcreator={LaTeX via pandoc}}

\title{Checking the Cumulative Exposure Model: Cumulative Chi-Squared
Goodness-of-Fit Tests for Interval-Monitored Step-Stress Life Tests}
\author{Nirian
Martín\\[4pt] \normalsize\itshape Department of Statistics and Operations Research and Interdisciplinary Mathematics Institute (IMI),\\ \normalsize\itshape Complutense University of Madrid, Spain}
\date{}

\begin{document}
\maketitle
\begin{abstract}
In an interval-monitored step-stress life test the data are the numbers
of devices that fail between consecutive inspections. These cells are
ordered in time, but the usual goodness-of-fit check, Pearson's
statistic, ignores the order. We propose goodness-of-fit tests based on
the cumulative residuals, the differences between the empirical and the
fitted distribution functions at the inspection times, with the
parameters estimated by the minimum density power divergence estimator.
The two statistics are Taguchi's cumulative chi-squared statistic and
its version with constant weights, the classical members of a family of
cumulative divergences for ordered categories, with the null
probabilities replaced by those of the fitted model. For the exponential
cumulative exposure model with a log-linear link we prove the existence,
uniqueness and consistency of the estimator under explicit conditions,
and we derive the limit distribution of the statistics under the null
hypothesis and under contiguous alternatives: a weighted chi-square
distribution with \(L-s\) terms, where \(L\) is the number of inspection
times and \(s\) the number of parameters. The critical values are
obtained by plugging the estimated weights into Imhof's formula, and the
procedure has asymptotic size \(\alpha\). A local-power analysis and a
simulation study show that the cumulative statistics are more powerful
than Pearson's statistic against a wrong lifetime distribution, and that
the limit distribution is accurate already with \(50\) devices. An
application to light bulbs tested under step voltage illustrates the
method.
\end{abstract}

\noindent\textit{Keywords:} accelerated life test; cumulative exposure model; density power divergence; goodness-of-fit; interval censoring; weighted chi-square distribution.

\hypertarget{introduction}{%
\section{Introduction}\label{introduction}}

Accelerated life tests shorten the time to failure of highly reliable
devices by testing them at stresses above the normal operating
conditions; the results are extrapolated to the normal stress through a
model that links the lifetime distribution to the stress (Nelson, 1990;
Escobar and Meeker, 2006). In a step-stress test the stress is raised at
prefixed times, and the cumulative exposure model (Sedyakin, 1966;
Nelson, 1980) relates the lifetime distribution under the varying stress
to the distributions under constant stresses. When the devices are not
monitored continuously but inspected at prefixed times, the data are the
numbers of failures between consecutive inspections, and inference is
based on a multinomial model whose cell probabilities depend on the
parameters of the lifetime distribution and of the link.

For this setting, robust estimation based on the density power
divergence (Basu et al., 1998) has been developed for exponential,
Weibull, gamma and lognormal lifetimes, together with Wald- and Rao-type
tests of hypotheses on the parameters; see the review of Balakrishnan,
Jaenada and Pardo (2026a), the thesis of Jaenada (2024) and
Balakrishnan, Jaenada and Pardo (2026b). All these procedures assume
that the parametric model is correct. Checking it is left to Pearson's
chi-square statistic, and the review lists goodness-of-fit testing for
interval-monitored step-stress data, possibly robust and based on the
minimum density power divergence estimator, as an open problem
(Balakrishnan, Jaenada and Pardo, 2026a, Open Problem 8).

In applications, the fitted models are checked with Pearson's statistic
computed with the maximum likelihood estimator. For a simple step-stress
test with \(n=100\) units, Balakrishnan, Jaenada and Pardo (2026a,
Section 12) report \(p\)-values between \(0.67\) and \(0.80\) for
exponential, Weibull, gamma and lognormal lifetimes, and conclude that
any of the four families could fit the data. Balakrishnan, Jaenada and
Pardo (2026b) use the same statistic to support Weibull lifetimes with
\(n=35\) and \(n=27\) devices. Two features of this practice motivate
the present work. First, the check uses the maximum likelihood estimator
even when the analysis relies on a robust estimator, and with the
minimum density power divergence estimator, \(\beta>0\), the limit of
Pearson's statistic is no longer a chi-square distribution but a
weighted chi-square distribution (Section 3.2). Second, Pearson's
statistic has little power to distinguish lifetime families that differ
smoothly over time, which is the typical form of misspecification in
life testing. Escobar and Meeker (2006, Section 10.2) stress that
accelerated test data usually have little power to detect any but
serious inadequacies of the assumed model, so the power of the check
matters.

Pearson's statistic, like any statistic built from the cells as an
unordered set, ignores that the cells of a life test are ordered in
time. A wrong lifetime distribution does not produce an isolated
discrepancy in one cell but a smooth, systematic one, which accumulates
over consecutive cells. Statistics based on the cumulative residuals,
the differences between the empirical and the fitted distribution
functions at the inspection times, are built to detect such departures.
For ordered categories with a known null distribution the idea is
classical: Taguchi's accumulation analysis adds the Pearson statistics
of the two-category tables obtained by collapsing the categories at each
cut point (Taguchi, 1974), and Hirotsu (1986) and Nair (1986, 1987)
studied this cumulative chi-squared statistic and its version with
constant weights; Hamada and Wu (1990) review its use in industrial
experiments. For discrete distributions with parameters estimated by
maximum likelihood, the quadratic cumulative statistics of Cramér--von
Mises and Anderson--Darling type were studied by Choulakian, Lockhart
and Stephens (1994), Spinelli (2001) and Lockhart, Spinelli and Stephens
(2007); Spinelli (2001) showed for the grouped exponential distribution
that they are not uniformly more powerful than Pearson's statistic.
Martín and de Frutos (2026) embed these statistics in a family of
cumulative and survival \(\phi\)-divergences and density power
divergences for ordered categories, the ordinal analogues of the
measures of Zografos (2023), with a common asymptotic theory for a
simple null hypothesis. The two statistics of this paper are the two
classical members of that family, Taguchi's statistic and the statistic
with constant weights, with the null probabilities replaced by those of
the fitted step-stress model.

The contributions of the paper are as follows. We propose
goodness-of-fit tests for the interval-monitored step-stress model based
on cumulative residuals, with the parameters estimated by the minimum
density power divergence estimator, of which the maximum likelihood
estimator is the case \(\beta=0\) (Section 3). For the exponential
cumulative exposure model with a log-linear link, we prove the
existence, uniqueness and consistency of the estimator under explicit
conditions, which the literature states without proof (Proposition 2),
and we derive the limit distribution of the statistics under the null
hypothesis and under contiguous alternatives (Theorems 1 and 2). We give
a plug-in calibration with asymptotic size \(\alpha\) (Corollary 1),
whose critical values are insensitive to the parameter in the designs
considered and which is accurate in finite samples (Section 5.1). A
local-power analysis locates the gain: the cumulative statistics are
more powerful than Pearson's statistic against a wrong lifetime
distribution, and react less to an excess of failures in one interior
cell (Section 4). A simulation study confirms the gain in finite samples
and describes the behaviour of the tests under contamination (Section
5), and an application to light bulbs tested under step voltage
completes the picture (Section 6).

The minimum density power divergence estimator limits the effect of
outlying counts on the fitted model, but not on the statistic: the
cumulative residuals compare the fitted distribution function with the
empirical one, and the empirical distribution function contains the
outlying counts unchanged. Goodness-of-fit tests whose size and power
are stable under contamination therefore require, in addition to a
robust estimator, a statistic in which cells with outlying counts
receive a reduced weight. The construction of such a statistic, to be
combined with the minimum density power divergence estimator, is the
subject of ongoing work (Section 7). Throughout, the design, that is,
the stress levels, the time of the change of stress and the inspection
times, is fixed and known; contamination of the design, which
Balakrishnan, Jaenada and Pardo (2024) consider for the estimator, is
outside the scope of this paper.

\hypertarget{model-and-estimation}{%
\section{Model and estimation}\label{model-and-estimation}}

\hypertarget{design-and-data}{%
\subsection{Design and data}\label{design-and-data}}

A sample of \(n\) devices is placed on test at stress \(x_{1}\); at the
prefixed time \(\tau_{1}\) the stress is raised to \(x_{2}>x_{1}\) and
kept until the end of the test \(\tau_{2}\). The devices are inspected
at prefixed times \(0=t_{0}<t_{1}<\cdots<t_{L}=\tau_{2}\), among which
is \(\tau_{1}=t_{m}\), \(1\leq m<L\). Let \(N_{j}\) be the number of
devices that fail in \((t_{j-1},t_{j}]\), \(j=1,\ldots,L\), and
\(N_{L+1}\) the number still working at \(t_{L}\);
\(\boldsymbol{N}=(N_{1},\ldots,N_{L+1})^{\top}\),
\(\widehat{\boldsymbol{p}}=\boldsymbol{N}/n\). Let \(\boldsymbol{C}\) be
the \(L\times(L+1)\) matrix with \(C(j,i)=1\) if \(i\leq j\) and \(0\)
otherwise; then
\(\widehat{\boldsymbol{F}}=\boldsymbol{C}\widehat{\boldsymbol{p}}\) is
the vector of empirical distribution functions at
\(t_{1},\ldots,t_{L}\).

If the lifetimes \(T_{1},\ldots,T_{n}\) are independent with common
distribution function \(G\), then \(\boldsymbol{N}\) has a multinomial
distribution \(\mathcal{M}(n,\boldsymbol{p})\) with
\(p_{j}=G(t_{j})-G(t_{j-1})\), \(j\leq L\), and \(p_{L+1}=1-G(t_{L})\).

\hypertarget{the-exponential-cumulative-exposure-model}{%
\subsection{The exponential cumulative exposure
model}\label{the-exponential-cumulative-exposure-model}}

Under the cumulative exposure model with exponential lifetimes of rate
\(\lambda_{i}\) at stress \(x_{i}\), the distribution function is
\(F(t)=1-\exp(-u(t))\), where the cumulative exposure \[
u(t)=\lambda_{1}\min(t,\tau_{1})+\lambda_{2}\max(t-\tau_{1},0)
\tag{1}
\] is the integral of the failure rate, which equals the rate of the
stress in force (Nelson, 1980; Kundu and Ganguly, 2017). The rates are
linked to the stress by
\(\lambda_{i}(\boldsymbol{\theta})=\theta_{0}\exp(\theta_{1}x_{i})\),
\(\boldsymbol{\theta}=(\theta_{0},\theta_{1})^{\top}\in\Theta=(0,\infty)\times\mathbb{R}\),
so \(s=2\), and we write \(F_{\boldsymbol{\theta}}\),
\(u(t;\boldsymbol{\theta})\), \(\boldsymbol{\pi}(\boldsymbol{\theta})\)
for the resulting distribution function, exposure and cell
probabilities,
\(\boldsymbol{F}_{\boldsymbol{\theta}}=\boldsymbol{C}\boldsymbol{\pi}(\boldsymbol{\theta})\),
and
\(\boldsymbol{J}(\boldsymbol{\theta})=\partial\boldsymbol{\pi}(\boldsymbol{\theta})/\partial\boldsymbol{\theta}^{\top}\).
The hypotheses are \[
H_{0}:\ \boldsymbol{p}=\boldsymbol{\pi}(\boldsymbol{\theta})\ \text{for some }\boldsymbol{\theta}\in\Theta
\qquad\text{versus}\qquad
H_{1}:\ \boldsymbol{p}\neq\boldsymbol{\pi}(\boldsymbol{\theta})\ \text{for every }\boldsymbol{\theta}\in\Theta,
\tag{2}
\] and \(\boldsymbol{\theta}^{\ast}\) denotes the true value under
\(H_{0}\). With two stress levels the map
\(\boldsymbol{\theta}\mapsto(\lambda_{1},\lambda_{2})\) is a bijection
of \(\Theta\) onto \((0,\infty)^{2}\), so the link imposes no
restriction: \(H_{0}\) states that the lifetimes are exponential at each
stress and combined by cumulative exposure. With exponential lifetimes
the cumulative exposure, tampered random variable and tampered failure
rate models coincide (Kundu and Ganguly, 2017, Section 1.5), so
\(H_{0}\) does not depend on that choice. The results of Sections 2 and
3 use the model only through the assumptions below and extend to any
smooth parametric model for \(\boldsymbol{\pi}(\boldsymbol{\theta})\)
that satisfies them, for instance the Weibull cumulative exposure model
or designs with \(k>2\) stress levels.

We use the following assumptions. Let
\(\boldsymbol{D}(\boldsymbol{\theta})=\operatorname{diag}(\boldsymbol{\pi}(\boldsymbol{\theta}))\)
and
\(\boldsymbol{\Sigma}(\boldsymbol{\theta})=\boldsymbol{D}(\boldsymbol{\theta})-\boldsymbol{\pi}(\boldsymbol{\theta})\boldsymbol{\pi}^{\top}(\boldsymbol{\theta})\).

\textbf{(A1)} \(\boldsymbol{\theta}^{\ast}\) is an interior point of the
open set \(\Theta\subseteq\mathbb{R}^{s}\).

\textbf{(A2)} \(\pi_{j}(\boldsymbol{\theta})>0\) for every \(j\) and
\(\boldsymbol{\theta}\in\Theta\).

\textbf{(A3)} \(\boldsymbol{\pi}(\boldsymbol{\theta})\) is twice
continuously differentiable on \(\Theta\), and
\(\boldsymbol{J}(\boldsymbol{\theta}^{\ast})\) has rank \(s\).

\textbf{(A4)} For every sequence
\((\boldsymbol{\theta}_{r})_{r\in\mathbb{N}}\) of points of \(\Theta\),
if
\(\boldsymbol{\pi}(\boldsymbol{\theta}_{r})\xrightarrow[r\to\infty]{}\boldsymbol{\pi}(\boldsymbol{\theta}^{\ast})\),
then
\(\boldsymbol{\theta}_{r}\xrightarrow[r\to\infty]{}\boldsymbol{\theta}^{\ast}\).

These are the conditions of Birch (1964).

\begin{remark}
This is the parametrisation of Jaenada (2024). Balakrishnan, Jaenada and Pardo (2026a) link the mean lifetime instead, $1/\lambda_{i}=e^{\theta_{0}+\theta_{1}x_{i}}$, and Balakrishnan, Jaenada and Pardo (2026b) write it as $e^{a_{0}+a_{1}x_{i}}$; their parameters correspond to $(-\log\theta_{0},-\theta_{1})$ in our notation. The rate parametrisation is kept here because the theory is stated in terms of the cell probabilities, which depend on $\boldsymbol{\theta}$ through the rates.
\end{remark}

\textbf{Lemma 1.} \emph{For the exponential cumulative exposure model,
(A1)--(A4) hold for every \(\boldsymbol{\theta}^{\ast}\in\Theta\),
provided \(1\leq m<L\), that is, at least one inspection time is at or
before \(\tau_{1}\) and one after it.}

The proofs are in the Appendix.

\hypertarget{the-minimum-density-power-divergence-estimator}{%
\subsection{The minimum density power divergence
estimator}\label{the-minimum-density-power-divergence-estimator}}

Let
\(\mathcal{P}_{L+1}=\{\boldsymbol{p}\in[0,1]^{L+1}:\boldsymbol{1}^{\top}\boldsymbol{p}=1\}\).
For \(\beta>0\) the density power divergence between
\(\boldsymbol{p},\boldsymbol{q}\in\mathcal{P}_{L+1}\) is \[
d_{\beta}(\boldsymbol{p},\boldsymbol{q})=\sum_{j=1}^{L+1}\Big(q_{j}^{1+\beta}-\Big(1+\frac{1}{\beta}\Big)p_{j}q_{j}^{\beta}+\frac{1}{\beta}p_{j}^{1+\beta}\Big),
\] and \(d_{0}\) is the Kullback--Leibler divergence. The minimum
density power divergence estimator
\(\widehat{\boldsymbol{\theta}}_{\beta}\) is any measurable minimiser of
\(d_{\beta}(\widehat{\boldsymbol{p}},\boldsymbol{\pi}(\boldsymbol{\theta}))\)
over \(\Theta\); \(\widehat{\boldsymbol{\theta}}_{0}\) is the maximum
likelihood estimator. An interior minimiser solves \[
\boldsymbol{\Psi}_{\beta}(\boldsymbol{\theta},\widehat{\boldsymbol{p}})\overset{\Delta}{=}\boldsymbol{J}^{\top}(\boldsymbol{\theta})\boldsymbol{D}^{\beta-1}(\boldsymbol{\theta})\big(\widehat{\boldsymbol{p}}-\boldsymbol{\pi}(\boldsymbol{\theta})\big)=\boldsymbol{0}.
\tag{3}
\] Let \[
\boldsymbol{A}_{\beta}=\boldsymbol{J}^{\top}(\boldsymbol{\theta}^{\ast})\boldsymbol{D}^{\beta-1}(\boldsymbol{\theta}^{\ast})\boldsymbol{J}(\boldsymbol{\theta}^{\ast}),\qquad
\boldsymbol{L}_{\beta}=\boldsymbol{A}_{\beta}^{-1}\boldsymbol{J}^{\top}(\boldsymbol{\theta}^{\ast})\boldsymbol{D}^{\beta-1}(\boldsymbol{\theta}^{\ast}),\qquad
\boldsymbol{M}_{\beta}=\boldsymbol{I}_{L+1}-\boldsymbol{J}(\boldsymbol{\theta}^{\ast})\boldsymbol{L}_{\beta}.
\]

\textbf{Proposition 1.} \emph{Assume (A1)--(A4) and let \(\beta\geq0\).
There are an open neighbourhood \(\mathcal{V}\) of
\(\boldsymbol{\pi}(\boldsymbol{\theta}^{\ast})\) in \(\mathbb{R}^{L+1}\)
and a continuously differentiable map
\(\boldsymbol{T}_{\beta}:\mathcal{V}\to\Theta\) with
\(\boldsymbol{T}_{\beta}(\boldsymbol{\pi}(\boldsymbol{\theta}^{\ast}))=\boldsymbol{\theta}^{\ast}\)
and
\(\partial\boldsymbol{T}_{\beta}(\boldsymbol{p})/\partial\boldsymbol{p}^{\top}=\boldsymbol{L}_{\beta}\)
at \(\boldsymbol{p}=\boldsymbol{\pi}(\boldsymbol{\theta}^{\ast})\), such
that for every \(\boldsymbol{p}\in\mathcal{V}\cap\mathcal{P}_{L+1}\),
\(\boldsymbol{T}_{\beta}(\boldsymbol{p})\) is the unique minimiser of
\(d_{\beta}(\boldsymbol{p},\boldsymbol{\pi}(\boldsymbol{\theta}))\) over
\(\Theta\).}

\textbf{Proposition 2.} \emph{Assume (A1)--(A4), let \(\beta\geq0\), and
let \(\widehat{\boldsymbol{p}}\) be the vector of proportions of a
sequence of multinomial vectors \(\mathcal{M}(n,\boldsymbol{p}_{n})\)
with} \[
\boldsymbol{p}_{n}\xrightarrow[n\to\infty]{}\boldsymbol{\pi}(\boldsymbol{\theta}^{\ast})
\] \emph{and
\(\sqrt{n}\|\widehat{\boldsymbol{p}}-\boldsymbol{\pi}(\boldsymbol{\theta}^{\ast})\|\)
bounded in probability. Then, with probability tending to one,
\(\widehat{\boldsymbol{\theta}}_{\beta}\) exists, is unique and equals
\(\boldsymbol{T}_{\beta}(\widehat{\boldsymbol{p}})\), and} \[
\sqrt{n}\big(\widehat{\boldsymbol{\theta}}_{\beta}-\boldsymbol{\theta}^{\ast}\big)=\boldsymbol{L}_{\beta}\sqrt{n}\big(\widehat{\boldsymbol{p}}-\boldsymbol{\pi}(\boldsymbol{\theta}^{\ast})\big)+\boldsymbol{R}_{n},\qquad\boldsymbol{R}_{n}\xrightarrow[n\to\infty]{P}\boldsymbol{0}.
\] \emph{If the null hypothesis (2) holds with true value
\(\boldsymbol{\theta}^{\ast}\), that is,} \[
\boldsymbol{p}_{n}=\boldsymbol{\pi}(\boldsymbol{\theta}^{\ast})\quad\text{for every }n,
\] \emph{then
\(\widehat{\boldsymbol{\theta}}_{\beta}\xrightarrow[n\to\infty]{\mathrm{a.s.}}\boldsymbol{\theta}^{\ast}\)
and} \[
\sqrt{n}\big(\widehat{\boldsymbol{\theta}}_{\beta}-\boldsymbol{\theta}^{\ast}\big)\xrightarrow[n\to\infty]{\mathcal{L}}\mathcal{N}_{s}\big(\boldsymbol{0},\boldsymbol{A}_{\beta}^{-1}\boldsymbol{K}_{\beta}\boldsymbol{A}_{\beta}^{-1}\big),\qquad
\boldsymbol{K}_{\beta}=\boldsymbol{J}^{\top}(\boldsymbol{\theta}^{\ast})\big(\boldsymbol{D}^{2\beta-1}(\boldsymbol{\theta}^{\ast})-\boldsymbol{\pi}^{\beta}(\boldsymbol{\theta}^{\ast})\boldsymbol{\pi}^{\beta\top}(\boldsymbol{\theta}^{\ast})\big)\boldsymbol{J}(\boldsymbol{\theta}^{\ast}).
\]

The limit under \(H_{0}\) is Theorem 2 of Balakrishnan, Jaenada and
Pardo (2026a), there stated without regularity conditions; the
existence, uniqueness and consistency are new. For \(\beta=0\), by
Proposition 3 of the Appendix the maximum likelihood estimator exists in
\(\Theta\), and is unique, if and only if \(N_{1}+\cdots+N_{m}\geq1\),
\(N_{1}<n\), \(N_{m+1}+\cdots+N_{L}\geq1\) and
\(N_{m+1}<n-\sum_{j\leq m}N_{j}\).

The tuning parameter trades efficiency for robustness of the fit. We use
the grid \(\beta\in\{0,0.2,0.4,0.6,0.8,1\}\) of Balakrishnan, Jaenada
and Pardo (2026a, 2026b) and a data-driven choice, the iterative version
of the criterion of Warwick and Jones (2005) proposed by Basak, Basu and
Jones (2021). For each \(\beta\) in a grid \(\mathcal{B}\subset[0,1]\)
it estimates the mean squared error of
\(\widehat{\boldsymbol{\theta}}_{\beta}\) by \[
\widehat{\mathrm{MSE}}(\beta;\beta_{\mathrm{P}})=\big\|\widehat{\boldsymbol{\eta}}_{\beta}-\widehat{\boldsymbol{\eta}}_{\beta_{\mathrm{P}}}\big\|^{2}+\frac{1}{n}\operatorname{tr}\big(\boldsymbol{A}_{\beta}^{-1}\boldsymbol{K}_{\beta}\boldsymbol{A}_{\beta}^{-1}\big)\big|_{\widehat{\boldsymbol{\eta}}_{\beta}},
\] where
\(\widehat{\boldsymbol{\eta}}_{\beta}=(\log\widehat{\lambda}_{1},\log\widehat{\lambda}_{2})^{\top}\)
is the estimator in the log-rates, in which the two components have
comparable scales, the covariance matrix of Proposition 2 is computed in
the same parametrisation, and \(\beta_{\mathrm{P}}\) is a pilot value.
Starting from \(\beta_{\mathrm{P}}=0.5\), the minimiser of
\(\widehat{\mathrm{MSE}}(\cdot;\beta_{\mathrm{P}})\) over
\(\mathcal{B}\) becomes the new pilot until it no longer changes; the
final value is \(\widehat{\beta}_{\mathrm{opt}}\). We take
\(\mathcal{B}=\{0,0.1,\ldots,1\}\). In the running example of Section
3.4 the efficiency of \(\widehat{\boldsymbol{\eta}}_{\beta}\) relative
to the maximum likelihood estimator, component by component, is
\((0.996,0.998)\) for \(\beta=0.2\), \((0.966,0.982)\) for \(\beta=0.6\)
and \((0.916,0.956)\) for \(\beta=1\).

\begin{remark}
The results of Sections 2 and 3 are proved for a fixed $\beta$; with $\widehat{\beta}_{\mathrm{opt}}$ the test uses the weights at $\widehat{\beta}_{\mathrm{opt}}$ as if this value were fixed. Since $\mathcal{B}$ is finite, if $\widehat{\beta}_{\mathrm{opt}}$ converged in probability to some $\beta_{0}\in\mathcal{B}$, then $\operatorname{P}(\widehat{\beta}_{\mathrm{opt}}=\beta_{0})\to1$ and Theorem 1 and Corollary 1 would apply with $\beta_{0}$. Under the model, however, both terms of $\widehat{\mathrm{MSE}}$ are of order $1/n$, so $\widehat{\beta}_{\mathrm{opt}}$ remains random in the limit, and the asymptotic size of the test with $\widehat{\beta}_{\mathrm{opt}}$ is not covered by our results. It is examined by simulation in Section 5.
\end{remark}

\hypertarget{test-statistics-and-their-asymptotic-distributions}{%
\section{Test statistics and their asymptotic
distributions}\label{test-statistics-and-their-asymptotic-distributions}}

\hypertarget{the-statistics}{%
\subsection{The statistics}\label{the-statistics}}

Let \(w:(0,1)\to(0,\infty)\) be continuous and
\(\boldsymbol{W}(\boldsymbol{\theta})=\operatorname{diag}(w(F_{\boldsymbol{\theta}}(t_{1})),\ldots,w(F_{\boldsymbol{\theta}}(t_{L})))\).
The statistic is the weighted sum of squared cumulative residuals \[
T_{w,\beta}=n\big(\widehat{\boldsymbol{F}}-\boldsymbol{F}_{\widehat{\boldsymbol{\theta}}_{\beta}}\big)^{\top}\boldsymbol{W}(\widehat{\boldsymbol{\theta}}_{\beta})\big(\widehat{\boldsymbol{F}}-\boldsymbol{F}_{\widehat{\boldsymbol{\theta}}_{\beta}}\big)=n\sum_{j=1}^{L}w\big(F_{\widehat{\boldsymbol{\theta}}_{\beta}}(t_{j})\big)\big(\widehat{F}_{j}-F_{\widehat{\boldsymbol{\theta}}_{\beta}}(t_{j})\big)^{2}.
\] We use the two weights of the classical members of the family of
Martín and de Frutos (2026). With \(w(F)=1/(F(1-F))\), \(T_{w,\beta}\)
is the member of their cumulative--survival statistic
\(T^{\mathrm{CS}}_{\phi}=2n(\phi''(1))^{-1}\mathcal{CSD}_{\phi}\) with
the Pearson generator \(\phi(t)=(t-1)^{2}\): the sum, over the
inspection times, of the Pearson statistics of the two-cell tables
``failed by \(t_{j}\)'' and ``working at \(t_{j}\)'', that is, Taguchi's
cumulative chi-squared statistic in its one-sample form, with the fitted
probabilities in the place of known ones. We denote it by
\(T^{\mathrm{CS}}\). With \(w(F)=1\), \(T_{w,\beta}\) is their density
power statistic \(T^{\mathrm{C}}_{a}=2n(1+a)^{-1}\mathcal{C}d_{a}\) with
\(a=1\), which is Nair's statistic with constant weights; we denote it
by \(T^{\mathrm{C}}_{1}\). For these two members the quadratic form is
the statistic itself, not an approximation. The tuning parameter
\(\beta\) enters only through the estimator. With \(\beta=0\) the two
statistics are close to the discrete Anderson--Darling and Cramér--von
Mises statistics \(A^{2}\) and \(W^{2}\) of Choulakian, Lockhart and
Stephens (1994) and Lockhart, Spinelli and Stephens (2007), which weight
the squared residual at \(t_{j}\) also by the cell probability \(p_{j}\)
(1994) or \((p_{j}+p_{j+1})/2\) (2007), so that they approximate
integrals with respect to \(F\); for a discrete uniform null
distribution the two conventions differ only by a constant factor
(Martín and de Frutos, 2026, Section 4.2). Here every inspection time
receives the same weight, as in the family of Martín and de Frutos
(2026): the inspection times are design points chosen by the
experimenter, not a partition of the support. Theorems 1 and 2 hold for
both weightings, with \(w\) replaced accordingly.

\begin{remark}
The other members of the family are not quadratic. Let $\phi$ be convex on $[0,\infty)$, twice continuously differentiable in a neighbourhood of $1$, with $\phi(1)=0$ and $\phi''(1)>0$, let $\phi^{\ast}(t)=\phi(t)-\phi'(1)(t-1)$ be its normalised generator, let $a>0$, and write $\widetilde{F}_{j}=F_{\widehat{\boldsymbol{\theta}}_{\beta}}(t_{j})$. The cumulative, survival and density power members are
$$
T^{\mathrm{C}}_{\phi}=\frac{2n}{\phi''(1)}\sum_{j=1}^{L}\widetilde{F}_{j}\,\phi^{\ast}\Big(\frac{\widehat{F}_{j}}{\widetilde{F}_{j}}\Big),\qquad
T^{\mathrm{S}}_{\phi}=\frac{2n}{\phi''(1)}\sum_{j=1}^{L}(1-\widetilde{F}_{j})\,\phi^{\ast}\Big(\frac{1-\widehat{F}_{j}}{1-\widetilde{F}_{j}}\Big),
$$
$T^{\mathrm{CS}}_{\phi}=T^{\mathrm{C}}_{\phi}+T^{\mathrm{S}}_{\phi}$, and $T^{\mathrm{C}}_{a}=2n(1+a)^{-1}\sum_{j}(\widetilde{F}_{j}^{1+a}-(1+a^{-1})\widetilde{F}_{j}^{a}\widehat{F}_{j}+a^{-1}\widehat{F}_{j}^{1+a})$ (Martín and de Frutos, 2026). Under (A1)--(A4), under $H_{0}$ and under the contiguous alternatives (4), each of them differs from $T_{w,\beta}$ with the weight $w(F)=1/F$, $1/(1-F)$, $1/(F(1-F))$ or $F^{a-1}$, respectively, by a term that tends to zero in probability (proof in the Appendix). Hence Theorems 1 and 2 and Corollary 1 apply to every member, with these weights. We use the two classical members only; the choice of the member is discussed by Martín and de Frutos (2026).
\end{remark}

For comparison we use Pearson's statistic
\(X^{2}_{\beta}=n(\widehat{\boldsymbol{p}}-\boldsymbol{\pi}(\widehat{\boldsymbol{\theta}}_{\beta}))^{\top}\boldsymbol{D}^{-1}(\widehat{\boldsymbol{\theta}}_{\beta})(\widehat{\boldsymbol{p}}-\boldsymbol{\pi}(\widehat{\boldsymbol{\theta}}_{\beta}))\).

\hypertarget{asymptotic-distribution-under-the-null-hypothesis}{%
\subsection{Asymptotic distribution under the null
hypothesis}\label{asymptotic-distribution-under-the-null-hypothesis}}

Let
\(\mathcal{S}_{0}=\{\boldsymbol{v}\in\mathbb{R}^{L+1}:\boldsymbol{1}^{\top}\boldsymbol{v}=0\}\)
and \[
\boldsymbol{\Gamma}_{\beta}=\boldsymbol{C}\boldsymbol{M}_{\beta}\boldsymbol{\Sigma}(\boldsymbol{\theta}^{\ast})\boldsymbol{M}_{\beta}^{\top}\boldsymbol{C}^{\top}.
\]

\textbf{Lemma 2.} \emph{Under (A2)--(A3): (i)
\(\operatorname{range}\boldsymbol{\Sigma}(\boldsymbol{\theta}^{\ast})=\mathcal{S}_{0}\)
and
\(\ker\boldsymbol{\Sigma}(\boldsymbol{\theta}^{\ast})=\operatorname{span}\{\boldsymbol{1}\}\);
(ii) \(\boldsymbol{M}_{\beta}\) is idempotent,
\(\boldsymbol{M}_{\beta}\boldsymbol{J}(\boldsymbol{\theta}^{\ast})=\boldsymbol{0}\)
and
\(\boldsymbol{1}^{\top}\boldsymbol{M}_{\beta}=\boldsymbol{1}^{\top}\);
\(\boldsymbol{M}_{\beta}\) is the projection along the column space of
\(\boldsymbol{J}(\boldsymbol{\theta}^{\ast})\) that is orthogonal in the
inner product with matrix
\(\boldsymbol{D}^{\beta-1}(\boldsymbol{\theta}^{\ast})\); (iii)
\(\operatorname{range}\boldsymbol{\Gamma}_{\beta}=\boldsymbol{C}\boldsymbol{M}_{\beta}(\mathcal{S}_{0})\)
and \(\operatorname{rank}\boldsymbol{\Gamma}_{\beta}=L-s\) for every
\(\beta\geq0\).}

\textbf{Theorem 1.} \emph{Under (A1)--(A4) and \(H_{0}\), for every
\(\beta\geq0\),} \[
T_{w,\beta}\xrightarrow[n\to\infty]{\mathcal{L}}\sum_{j=1}^{L-s}\lambda_{j}Z_{j}^{2},
\] \emph{where \(Z_{1},\ldots,Z_{L-s}\) are independent
\(\mathcal{N}(0,1)\) variables and
\(\lambda_{1}\geq\cdots\geq\lambda_{L-s}>0\) are the nonzero eigenvalues
of
\(\boldsymbol{\Gamma}_{\beta}\boldsymbol{W}(\boldsymbol{\theta}^{\ast})\).
The limit is a chi-square distribution if and only if
\(\boldsymbol{\Gamma}_{\beta}\boldsymbol{W}(\boldsymbol{\theta}^{\ast})\boldsymbol{\Gamma}_{\beta}=\boldsymbol{\Gamma}_{\beta}\).}

For \(\beta=0\), Theorem 1 is the result of Lockhart, Spinelli and
Stephens (2007), with \(\boldsymbol{\Gamma}_{0}\) their
\(\boldsymbol{\Sigma}_{u}\). The same argument gives the limit of
\(X^{2}_{\beta}\): a weighted chi-square distribution with the
eigenvalues of
\(\boldsymbol{M}_{\beta}\boldsymbol{\Sigma}(\boldsymbol{\theta}^{\ast})\boldsymbol{M}_{\beta}^{\top}\boldsymbol{D}^{-1}(\boldsymbol{\theta}^{\ast})\),
which are \(L-s\) ones for \(\beta=0\) but not for \(\beta>0\). Weighted
chi-square limits with estimated weights also appear for
divergence-based tests of parametric hypotheses (Basu et al., 2018;
Jaenada, 2024, Theorem 4.7.4).

\hypertarget{asymptotic-distribution-under-contiguous-alternatives}{%
\subsection{Asymptotic distribution under contiguous
alternatives}\label{asymptotic-distribution-under-contiguous-alternatives}}

\textbf{Theorem 2.} \emph{Assume (A1)--(A4), let
\(\boldsymbol{d}\in\mathcal{S}_{0}\), and let
\(\boldsymbol{N}\sim\mathcal{M}(n,\boldsymbol{p}_{n})\) with the
contiguous alternatives} \[
\boldsymbol{p}_{n}=\boldsymbol{\pi}(\boldsymbol{\theta}^{\ast})+\frac{\boldsymbol{d}}{\sqrt{n}}.
\tag{4}
\] \emph{Write
\(\boldsymbol{W}^{1/2}(\boldsymbol{\theta}^{\ast})\boldsymbol{\Gamma}_{\beta}\boldsymbol{W}^{1/2}(\boldsymbol{\theta}^{\ast})=\sum_{j=1}^{L-s}\lambda_{j}\boldsymbol{q}_{j}\boldsymbol{q}_{j}^{\top}\)
with orthonormal \(\boldsymbol{q}_{j}\). Then} \[
T_{w,\beta}\xrightarrow[n\to\infty]{\mathcal{L}}\sum_{j=1}^{L-s}\lambda_{j}(Z_{j}+\delta_{j})^{2},\qquad\delta_{j}=\lambda_{j}^{-1/2}\boldsymbol{q}_{j}^{\top}\boldsymbol{W}^{1/2}(\boldsymbol{\theta}^{\ast})\boldsymbol{C}\boldsymbol{M}_{\beta}\boldsymbol{d}.
\] \emph{In particular, a departure in the column space of
\(\boldsymbol{J}(\boldsymbol{\theta}^{\ast})\), a change of the
parameter within the model, has \(\boldsymbol{\delta}=\boldsymbol{0}\).}

\hypertarget{calibration}{%
\subsection{Calibration}\label{calibration}}

The weights depend on the unknown \(\boldsymbol{\theta}^{\ast}\). Let
\(\widehat{\lambda}_{j}\) be the nonzero eigenvalues of
\(\widehat{\boldsymbol{\Gamma}}_{\beta}\boldsymbol{W}(\widehat{\boldsymbol{\theta}}_{\beta})\),
where \(\widehat{\boldsymbol{\Gamma}}_{\beta}\) is
\(\boldsymbol{\Gamma}_{\beta}\) evaluated at
\(\widehat{\boldsymbol{\theta}}_{\beta}\), and let
\(\widehat{q}_{1-\alpha}\) be the \(1-\alpha\) quantile of
\(\sum_{j}\widehat{\lambda}_{j}Z_{j}^{2}\), computed by Imhof's (1961)
formula.

\textbf{Corollary 1.} \emph{Assume (A1)--(A4) and let \(q_{1-\alpha}\)
be the \(1-\alpha\) quantile of the limit of Theorem 1. Under
\(H_{0}\),} \[
\operatorname{P}\big(T_{w,\beta}>\widehat{q}_{1-\alpha}\big)\xrightarrow[n\to\infty]{}\alpha,
\] \emph{and under the contiguous alternatives (4),} \[
\operatorname{P}\big(T_{w,\beta}>\widehat{q}_{1-\alpha}\big)\xrightarrow[n\to\infty]{}\operatorname{P}\Big\{\sum_{j=1}^{L-s}\lambda_{j}(Z_{j}+\delta_{j})^{2}>q_{1-\alpha}\Big\}.
\]

All the tests in this paper are calibrated in this way, with the limit
distribution evaluated exactly by Imhof's formula rather than
approximated, so that differences between statistics are not differences
of calibration. The word exact refers here to the evaluation of the
limit; how well the limit approximates the distribution of the statistic
for finite \(n\) is examined in Section 5.1. Approximations by moments
are not recommended. The two-moment approximation \(c\chi^{2}_{\nu}\),
with
\(c=\sum_{j}\widehat{\lambda}_{j}^{2}/\sum_{j}\widehat{\lambda}_{j}\)
and
\(\nu=(\sum_{j}\widehat{\lambda}_{j})^{2}/\sum_{j}\widehat{\lambda}_{j}^{2}\),
has nearly the right size at level \(0.05\) in the designs of this paper
(between \(0.0495\) and \(0.0511\)), but it is liberal further in the
tail: at nominal level \(0.01\) its size is between \(0.0125\) and
\(0.0143\), and at \(0.001\) between \(0.0019\) and \(0.0028\). In these
designs the weighted chi-square limit has a heavier right tail than the
\(c\chi^{2}_{\nu}\) with the same mean and variance, so the approximate
\(p\)-values are too small. Matching three cumulants with
\(a+b\chi^{2}_{d}\) (Imhof, 1961) is accurate at \(0.01\) in these
designs but still errs at \(0.001\).

Table 1 gives the weights for the design of Section 2.9 of Jaenada
(2024): \(x_{1}=35\), \(x_{2}=45\), \(\tau_{1}=25\), inspections at
\(10,15,20,25,30,35,40,45,50,60,70\) (\(L=11\)),
\(\boldsymbol{\theta}^{\ast}=(0.003,0.03)^{\top}\). Three features
matter. First, estimating \(\boldsymbol{\theta}\) reduces the critical
value of \(T^{\mathrm{CS}}\) from \(32.94\) to about \(6.2\), so the
critical values for a known parameter cannot be used. Second, the
critical value grows with \(\beta\), from \(6.17\) to \(7.04\) for
\(T^{\mathrm{CS}}\) and from \(1.04\) to \(1.37\) for
\(T^{\mathrm{C}}_{1}\) as \(\beta\) goes from \(0\) to \(1\), whereas
that of \(X^{2}\) hardly moves. Using the value for \(\beta=0\) after a
fit with \(\beta\) gives asymptotic sizes \(0.053\), \(0.058\),
\(0.064\), \(0.068\) and \(0.073\) for \(T^{\mathrm{CS}}\) and
\(0.063\), \(0.075\), \(0.086\), \(0.094\) and \(0.101\) for
\(T^{\mathrm{C}}_{1}\), for \(\beta=0.2,0.4,0.6,0.8,1\): the weights
must match the estimator. Third, over
\(\theta_{0}\in[0.8,1.2]\times0.003\) and
\(\theta_{1}\in[0.9,1.1]\times0.03\), the critical value of
\(T^{\mathrm{CS}}\) stays in \([6.12,6.24]\) for \(\beta=0\) and
\([6.22,6.34]\) for \(\beta=0.2\): the plug-in is insensitive to the
estimation error.

\begin{table}[htbp]
\centering
\caption{Four largest weights of the limit distribution, their sum and the $95\%$ quantile $q_{0.95}$ of the limit (Imhof's formula), running example of Section 3.4.}
\smallskip
\small
\setlength{\tabcolsep}{4.5pt}
\begin{tabular}{llcrrrrrr}
\toprule
Statistic & $\boldsymbol{\theta}$ & $\beta$ & $\lambda_{1}$ & $\lambda_{2}$ & $\lambda_{3}$ & $\lambda_{4}$ & $\sum_{j}\lambda_{j}$ & $q_{0.95}$ \\
\midrule
$T^{\mathrm{CS}}$ & known & -- & 7.598 & 1.670 & 0.637 & 0.355 & 11.000 & 32.94 \\
 & estimated & 0 & 0.947 & 0.774 & 0.260 & 0.210 & 2.619 & 6.17 \\
 & estimated & 0.2 & 1.065 & 0.661 & 0.272 & 0.222 & 2.646 & 6.28 \\
 & estimated & 0.4 & 1.176 & 0.573 & 0.284 & 0.234 & 2.694 & 6.47 \\
 & estimated & 0.6 & 1.264 & 0.511 & 0.297 & 0.243 & 2.741 & 6.67 \\
 & estimated & 0.8 & 1.331 & 0.474 & 0.309 & 0.248 & 2.787 & 6.85 \\
 & estimated & 1 & 1.384 & 0.467 & 0.311 & 0.250 & 2.837 & 7.04 \\
\midrule
$T^{\mathrm{C}}_{1}$ & known & -- & 1.482 & 0.265 & 0.109 & 0.056 & 2.039 & 6.29 \\
 & estimated & 0 & 0.196 & 0.075 & 0.056 & 0.028 & 0.433 & 1.04 \\
 & estimated & 0.2 & 0.223 & 0.065 & 0.058 & 0.029 & 0.453 & 1.12 \\
 & estimated & 0.4 & 0.247 & 0.064 & 0.054 & 0.030 & 0.474 & 1.21 \\
 & estimated & 0.6 & 0.266 & 0.067 & 0.049 & 0.031 & 0.492 & 1.27 \\
 & estimated & 0.8 & 0.279 & 0.071 & 0.045 & 0.033 & 0.507 & 1.33 \\
 & estimated & 1 & 0.289 & 0.077 & 0.042 & 0.034 & 0.521 & 1.37 \\
\midrule
$X^{2}$ & estimated & 0 & 1.000 & 1.000 & 1.000 & 1.000 & 9.000 & 16.92 \\
 & estimated & 0.2 & 1.005 & 1.002 & 1.000 & 1.000 & 9.007 & 16.93 \\
 & estimated & 0.4 & 1.019 & 1.006 & 1.000 & 1.000 & 9.026 & 16.97 \\
 & estimated & 0.6 & 1.044 & 1.010 & 1.000 & 1.000 & 9.054 & 17.02 \\
 & estimated & 0.8 & 1.079 & 1.013 & 1.000 & 1.000 & 9.092 & 17.09 \\
 & estimated & 1 & 1.123 & 1.014 & 1.000 & 1.000 & 9.138 & 17.18 \\
\bottomrule
\end{tabular}
\end{table}

\hypertarget{consistency}{%
\subsection{Consistency}\label{consistency}}

\textbf{Proposition 4.} \emph{Let
\(\boldsymbol{N}\sim\mathcal{M}(n,\boldsymbol{p})\) with
\(\boldsymbol{p}\) in the alternative hypothesis \(H_{1}\) of (2), and
suppose that \(\widehat{\boldsymbol{\theta}}_{\beta}\) converges in
probability to a point
\(\boldsymbol{\theta}_{\beta}(\boldsymbol{p})\in\Theta\). Then
\(T_{w,\beta}\xrightarrow[n\to\infty]{P}\infty\), and the test of
Corollary 1 has power tending to one.}

The assumption holds, for instance, when
\(d_{\beta}(\boldsymbol{p},\boldsymbol{\pi}(\boldsymbol{\theta}))\) has
a unique minimiser in \(\Theta\) that is well separated, by the argument
that gives (A.2) in the proof of Proposition 1.

\hypertarget{where-the-cumulative-statistics-gain}{%
\section{Where the cumulative statistics
gain}\label{where-the-cumulative-statistics-gain}}

Theorem 2 gives the asymptotic power against any sequence of contiguous
alternatives. We consider three kinds of departure, at level \(0.05\),
in two designs: the design of Table 1, and the light-bulb design of
Section 6 (grid A) at the fitted rates.

The first departure is a Weibull lifetime distribution,
\(F(t)=1-\exp(-u(t)^{\eta})\) with \(\eta=1+h/\sqrt{n}\) and the rates
fixed, which is the Weibull cumulative exposure model of Balakrishnan,
Jaenada and Pardo (2026b); then
\(\boldsymbol{d}=h\,\partial\boldsymbol{\pi}/\partial\eta\) at
\(\eta=1\), with
\(\partial F(t_{j})/\partial\eta=e^{-u_{j}}u_{j}\log u_{j}\). The second
is a gamma lifetime distribution, \(F(t)=P_{k}(u(t))\), with \(P_{k}\)
the gamma distribution function with shape \(k=1+h/\sqrt{n}\) and mean
one. The third is an anomalous count: a proportion \(h/\sqrt{n}\) of
devices fail in the third interval whatever their lifetime, so that
\(\boldsymbol{d}=h(\boldsymbol{e}_{3}-\boldsymbol{\pi}(\boldsymbol{\theta}^{\ast}))\),
which is the contamination scheme of Jaenada (2024, Section 2.9). The
first two are smooth departures: a failure rate that increases with age.
The third is concentrated in one cell.

\begin{table}[htbp]
\centering
\caption{Asymptotic rejection probabilities at level $0.05$ (Theorem 2), computed from $2\times10^{5}$ draws of the limit; Monte Carlo standard error at most $0.0011$.}
\smallskip
\small
\begin{tabular}{llc rrr c rrr}
\toprule
 & & & \multicolumn{3}{c}{Running example} & & \multicolumn{3}{c}{Light bulbs, grid A} \\
\cmidrule{4-6}\cmidrule{8-10}
Departure & $\beta$ & & $T^{\mathrm{CS}}$ & $T^{\mathrm{C}}_{1}$ & $X^{2}$ & & $T^{\mathrm{CS}}$ & $T^{\mathrm{C}}_{1}$ & $X^{2}$ \\
\midrule
Weibull, $h=4$ & 0 & & 0.234 & 0.150 & 0.148 & & 0.780 & 0.721 & 0.452 \\
 & 0.2 & & 0.192 & 0.114 & 0.147 & & 0.781 & 0.723 & 0.454 \\
 & 0.4 & & 0.159 & 0.092 & 0.148 & & 0.780 & 0.719 & 0.452 \\
 & 0.6 & & 0.134 & 0.081 & 0.152 & & 0.781 & 0.714 & 0.454 \\
 & 0.8 & & 0.115 & 0.075 & 0.154 & & 0.779 & 0.702 & 0.454 \\
 & 1 & & 0.103 & 0.073 & 0.158 & & 0.772 & 0.687 & 0.453 \\
\midrule
Gamma, $h=4$ & 0 & & 0.160 & 0.097 & 0.105 & & 0.518 & 0.448 & 0.243 \\
 & 0.2 & & 0.132 & 0.077 & 0.105 & & 0.520 & 0.446 & 0.242 \\
 & 0.4 & & 0.110 & 0.067 & 0.106 & & 0.517 & 0.438 & 0.241 \\
 & 0.6 & & 0.097 & 0.063 & 0.107 & & 0.518 & 0.435 & 0.241 \\
 & 0.8 & & 0.088 & 0.062 & 0.109 & & 0.512 & 0.423 & 0.245 \\
 & 1 & & 0.084 & 0.061 & 0.114 & & 0.502 & 0.408 & 0.245 \\
\midrule
Anomalous count, $h=1$ & 0 & & 0.707 & 0.435 & 0.936 & & 0.252 & 0.245 & 0.623 \\
 & 0.2 & & 0.681 & 0.375 & 0.936 & & 0.256 & 0.249 & 0.620 \\
 & 0.4 & & 0.648 & 0.332 & 0.937 & & 0.262 & 0.251 & 0.623 \\
 & 0.6 & & 0.624 & 0.318 & 0.938 & & 0.260 & 0.247 & 0.620 \\
 & 0.8 & & 0.605 & 0.319 & 0.939 & & 0.260 & 0.242 & 0.621 \\
 & 1 & & 0.593 & 0.335 & 0.937 & & 0.255 & 0.233 & 0.623 \\
\bottomrule
\end{tabular}
\end{table}

Table 2 shows a consistent pattern. Against a wrong lifetime
distribution, \(T^{\mathrm{CS}}\) with \(\beta=0\) is more powerful than
\(X^{2}\) in both designs, by a factor of up to \(2.1\) in the
light-bulb design. A smooth departure moves all the cumulative residuals
in the same direction, and the cumulative statistics concentrate their
weights on such directions (the first two weights of \(T^{\mathrm{CS}}\)
in Table 1 carry \(66\%\) of the total), whereas \(X^{2}\) spreads its
power evenly over \(L-s\) directions. Against an excess of failures in
the third interval, the cumulative statistics react less than \(X^{2}\):
a spike in one cell enters all later partial sums with the same small
weight, and the fit absorbs part of it. This is a consequence of
cumulating, not a robustness property: Section 5.3 shows that it depends
on the position of the cell. The effect of \(\beta\) depends on the
design. In the light-bulb design it is negligible: the local power of
every statistic changes by at most \(0.04\) over the grid. In the
running example the power of the cumulative statistics against the
smooth departures decreases steadily with \(\beta\), that of
\(T^{\mathrm{CS}}\) against the Weibull departure from \(0.234\) for
\(\beta=0\) to \(0.192\), \(0.134\) and \(0.103\) for \(\beta=0.2\),
\(0.6\) and \(1\), while that of \(X^{2}\) increases slightly; for
\(\beta\geq0.6\), \(X^{2}\) is the more powerful of the two. In that
design the two stress levels give similar rates, and the fit with large
\(\beta\), which downweights the cells of small probability, absorbs
more of a smooth departure into the parameters. Against the anomalous
count, \(\beta\) lowers the rejection probability of the cumulative
statistics only moderately. In the absence of contamination, the gain of
the cumulative statistics is therefore largest with \(\beta\) close to
\(0\), and the choice of \(\beta\) matters for the power of the test,
not only for the robustness of the fit; Section 5.3 shows the other
side, with contamination of an early cell, where \(\beta>0\) increases
the power.

No statistic is uniformly best. Spinelli (2001) showed that cumulative
statistics may lose against Pearson's statistic for departures that are
local in time, and the same holds here for departures concentrated
around the change of stress. A test directed at a specific alternative,
such as the Wald-type test of \(\eta=1\) in the Weibull model
(Balakrishnan, Jaenada and Pardo, 2026b), is more powerful against that
alternative; the cumulative statistics need no specified alternative.

\hypertarget{simulation-study}{%
\section{Simulation study}\label{simulation-study}}

We simulate the running example of Section 3.4. The estimators have
\(\beta\in\{0,0.2,0.4,0.6,0.8,1\}\), and we also use the data-driven
\(\widehat{\beta}_{\mathrm{opt}}\) of Section 2.3, chosen in each
replication. Each replication fits
\(\widehat{\boldsymbol{\theta}}_{\beta}\) numerically, computes the
weights at \(\widehat{\boldsymbol{\theta}}_{\beta}\), and computes the
\(p\)-value of each statistic, \(X^{2}\) included, from the limit of
Theorem 1 evaluated by Imhof's formula; the same calibration is used for
every statistic and every \(\beta\); with
\(\widehat{\beta}_{\mathrm{opt}}\) the weights are those of the selected
value. The maximum likelihood estimator existed in all replications with
\(n\geq100\) and failed to exist in one of the \(2\times10^{4}\) samples
with \(n=50\) (Proposition 3).

\hypertarget{accuracy-of-the-asymptotic-distribution}{%
\subsection{Accuracy of the asymptotic
distribution}\label{accuracy-of-the-asymptotic-distribution}}

Table 3 compares, under the null model, the quantiles of the limit of
Theorem 1 at \(\boldsymbol{\theta}^{\ast}\) with the empirical quantiles
of the statistics over \(10^{4}\) samples. The differences have no trend
in \(n\) and are of the size of the Monte Carlo error. Equivalently, the
proportion of samples in which the statistic exceeds the quantile of the
limit lies between \(0.043\) and \(0.056\) at level \(0.05\), and
between \(0.0085\) and \(0.0129\) at level \(0.01\), for every
statistic, every \(\beta\) and every \(n\in\{50,100,200,500\}\); the
Monte Carlo standard errors are \(0.0022\) and \(0.0010\). The limit is
therefore accurate already with \(n=50\), which is the order of the
sample sizes of the applications cited in Section 1. Table 3 uses the
true parameter and \(\beta\in\{0,0.2\}\); the effect of plugging in
\(\widehat{\boldsymbol{\theta}}_{\beta}\), for every \(\beta\) of the
grid, is included in Table 4.

\begin{table}[htbp]
\centering
\caption{Accuracy of the limit distribution under the null model, design of Table 1: quantiles $q_{1-\alpha}$ of the limit of Theorem 1 at $\boldsymbol{\theta}^{\ast}$ (Imhof's formula) and empirical quantiles of the statistic over $10^{4}$ samples of size $n$.}
\smallskip
\small
\setlength{\tabcolsep}{4pt}
\begin{tabular}{lc c r rrrr c r rrrr}
\toprule
 & & & \multicolumn{5}{c}{$\alpha=0.05$} & & \multicolumn{5}{c}{$\alpha=0.01$} \\
\cmidrule{4-8}\cmidrule{10-14}
 & & & & \multicolumn{4}{c}{Empirical, $n=$} & & & \multicolumn{4}{c}{Empirical, $n=$} \\
\cmidrule{5-8}\cmidrule{11-14}
Statistic & $\beta$ & & Limit & 50 & 100 & 200 & 500 & & Limit & 50 & 100 & 200 & 500 \\
\midrule
$T^{\mathrm{CS}}$ & 0 & & 6.17 & 6.02 & 6.17 & 6.36 & 6.10 & & 8.98 & 8.85 & 8.83 & 9.31 & 8.80 \\
 & 0.2 & & 6.28 & 6.48 & 6.38 & 6.17 & 6.29 & & 9.26 & 9.32 & 9.58 & 9.06 & 9.13 \\
\midrule
$T^{\mathrm{C}}_{1}$ & 0 & & 1.04 & 1.03 & 1.05 & 1.06 & 1.02 & & 1.58 & 1.55 & 1.57 & 1.68 & 1.55 \\
 & 0.2 & & 1.12 & 1.14 & 1.13 & 1.11 & 1.11 & & 1.74 & 1.75 & 1.76 & 1.72 & 1.73 \\
\midrule
$X^{2}$ & 0 & & 16.92 & 16.62 & 16.86 & 16.96 & 17.04 & & 21.67 & 21.98 & 21.38 & 21.85 & 21.94 \\
 & 0.2 & & 16.93 & 16.61 & 16.80 & 16.83 & 16.76 & & 21.68 & 21.59 & 21.87 & 21.69 & 21.61 \\
\bottomrule
\end{tabular}
\end{table}

\hypertarget{size-and-power}{%
\subsection{Size and power}\label{size-and-power}}

Table 4 gives the rejection rates of the plug-in tests of Corollary 1,
with \(n\in\{100,200\}\) and \(5000\) replications, so that the Monte
Carlo standard error is about \(0.003\) for the size and at most
\(0.007\) for the power. The data are generated from the null model at
\(\boldsymbol{\theta}^{\ast}\); from Weibull lifetimes,
\(F(t)=1-\exp(-u(t;\boldsymbol{\theta}^{\ast})^{\eta})\),
\(\eta\in\{1.2,1.4\}\); and from gamma lifetimes with shape \(1.5\) and
mean one in the scale of \(u\),
\(F(t)=P_{1.5}(u(t;\boldsymbol{\theta}^{\ast}))\).

\begin{table}[htbp]
\centering
\caption{Empirical rejection rates at level $0.05$ with the plug-in calibration of Corollary 1; running example, $5000$ replications.}
\smallskip
\small
\setlength{\tabcolsep}{4.5pt}
\begin{tabular}{ll rrrrrr r}
\toprule
 & & \multicolumn{6}{c}{$\beta$} & \\
\cmidrule{3-8}
Data & Statistic & 0 & 0.2 & 0.4 & 0.6 & 0.8 & 1 & $\widehat{\beta}_{\mathrm{opt}}$ \\
\midrule
Null model, $n=100$ & $T^{\mathrm{CS}}$ & 0.048 & 0.050 & 0.050 & 0.051 & 0.052 & 0.051 & 0.049 \\
 & $T^{\mathrm{C}}_{1}$ & 0.052 & 0.051 & 0.052 & 0.052 & 0.051 & 0.050 & 0.055 \\
 & $X^{2}$ & 0.050 & 0.052 & 0.052 & 0.052 & 0.053 & 0.054 & 0.050 \\
\midrule
Null model, $n=200$ & $T^{\mathrm{CS}}$ & 0.055 & 0.054 & 0.055 & 0.056 & 0.056 & 0.057 & 0.051 \\
 & $T^{\mathrm{C}}_{1}$ & 0.054 & 0.056 & 0.057 & 0.058 & 0.058 & 0.058 & 0.057 \\
 & $X^{2}$ & 0.052 & 0.052 & 0.052 & 0.052 & 0.052 & 0.053 & 0.052 \\
\midrule
Weibull, $\eta=1.2$, $n=100$ & $T^{\mathrm{CS}}$ & 0.074 & 0.072 & 0.066 & 0.062 & 0.057 & 0.056 & 0.079 \\
 & $T^{\mathrm{C}}_{1}$ & 0.069 & 0.063 & 0.059 & 0.056 & 0.054 & 0.053 & 0.071 \\
 & $X^{2}$ & 0.071 & 0.073 & 0.074 & 0.074 & 0.075 & 0.076 & 0.071 \\
\midrule
Weibull, $\eta=1.2$, $n=200$ & $T^{\mathrm{CS}}$ & 0.124 & 0.109 & 0.095 & 0.086 & 0.078 & 0.074 & 0.127 \\
 & $T^{\mathrm{C}}_{1}$ & 0.096 & 0.081 & 0.071 & 0.069 & 0.066 & 0.063 & 0.097 \\
 & $X^{2}$ & 0.096 & 0.096 & 0.097 & 0.099 & 0.100 & 0.103 & 0.095 \\
\midrule
Weibull, $\eta=1.4$, $n=100$ & $T^{\mathrm{CS}}$ & 0.148 & 0.133 & 0.118 & 0.106 & 0.095 & 0.091 & 0.154 \\
 & $T^{\mathrm{C}}_{1}$ & 0.131 & 0.112 & 0.100 & 0.093 & 0.088 & 0.085 & 0.134 \\
 & $X^{2}$ & 0.110 & 0.113 & 0.116 & 0.122 & 0.125 & 0.129 & 0.109 \\
\midrule
Weibull, $\eta=1.4$, $n=200$ & $T^{\mathrm{CS}}$ & 0.275 & 0.243 & 0.214 & 0.191 & 0.173 & 0.156 & 0.278 \\
 & $T^{\mathrm{C}}_{1}$ & 0.218 & 0.188 & 0.165 & 0.146 & 0.136 & 0.129 & 0.219 \\
 & $X^{2}$ & 0.178 & 0.183 & 0.187 & 0.191 & 0.194 & 0.199 & 0.178 \\
\midrule
Gamma, shape $1.5$, $n=100$ & $T^{\mathrm{CS}}$ & 0.140 & 0.120 & 0.105 & 0.095 & 0.088 & 0.082 & 0.144 \\
 & $T^{\mathrm{C}}_{1}$ & 0.099 & 0.085 & 0.076 & 0.069 & 0.067 & 0.063 & 0.100 \\
 & $X^{2}$ & 0.109 & 0.113 & 0.118 & 0.123 & 0.129 & 0.133 & 0.109 \\
\midrule
Gamma, shape $1.5$, $n=200$ & $T^{\mathrm{CS}}$ & 0.228 & 0.192 & 0.163 & 0.139 & 0.118 & 0.108 & 0.229 \\
 & $T^{\mathrm{C}}_{1}$ & 0.140 & 0.114 & 0.098 & 0.087 & 0.081 & 0.077 & 0.141 \\
 & $X^{2}$ & 0.145 & 0.151 & 0.157 & 0.161 & 0.169 & 0.176 & 0.145 \\
\bottomrule
\end{tabular}
\end{table}

The plug-in tests hold their level for every \(\beta\) and for
\(\widehat{\beta}_{\mathrm{opt}}\): the empirical sizes lie between
\(0.048\) and \(0.058\), and the largest, for \(T^{\mathrm{C}}_{1}\)
with \(n=200\) and \(\beta\geq0.6\), exceeds \(0.05\) by about \(2.5\)
Monte Carlo standard errors. Against Weibull and gamma lifetimes,
\(T^{\mathrm{CS}}\) with \(\beta=0\) is the most powerful test, with a
gain over \(X^{2}\) of \(54\%\) at \(\eta=1.4\) and of \(57\%\) against
the gamma lifetimes, both with \(n=200\). As in the local analysis of
Section 4, the power of the cumulative statistics decreases steadily
with \(\beta\), while that of \(X^{2}\) increases slightly: with
\(n=200\), \(T^{\mathrm{CS}}\) and \(X^{2}\) have the same power at
\(\eta=1.4\) for \(\beta=0.6\), and \(X^{2}\) is more powerful for
\(\beta\geq0.8\), and for \(\beta\geq0.6\) against the gamma lifetimes.
The data-driven \(\widehat{\beta}_{\mathrm{opt}}\) selects \(\beta=0\)
in \(56\%\) to \(58\%\) of the samples under the null model and in
\(72\%\) to \(90\%\) under the smooth departures, with mean between
\(0.02\) and \(0.09\), and the tests with
\(\widehat{\beta}_{\mathrm{opt}}\) have the power of the tests with
\(\beta=0\); under contamination of the first interval the criterion
selects larger values of \(\beta\) (Section 5.3). In this design the two
stress levels give similar failure rates
(\(\lambda_{2}/\lambda_{1}=1.35\)), so the step-stress structure plays a
small role; a design with a larger acceleration, such as the light-bulb
design of Section 6, gives larger gains and is insensitive to \(\beta\)
(Table 2).

\hypertarget{contamination}{%
\subsection{Contamination}\label{contamination}}

Table 5 moves a proportion \(\varepsilon\in\{0.05,0.10\}\) of the
devices to one cell,
\(\boldsymbol{p}=(1-\varepsilon)\boldsymbol{p}_{0}+\varepsilon\boldsymbol{e}_{j}\),
where \(\boldsymbol{p}_{0}\) is the null model or the Weibull model with
\(\eta=1.4\), and the cell is the first interval, which determines the
estimate of \(\lambda_{1}\), the third interval, as in Jaenada (2024,
Section 2.9), or the survivors. A goodness-of-fit test whose size and
power were stable under contamination would reject close to \(0.05\) in
the first block and close to the uncontaminated power in the second.

\begin{table}[htbp]
\centering
\caption{Empirical rejection rates at level $0.05$ when a proportion $\varepsilon$ of the devices is moved to one cell, $\boldsymbol{p}=(1-\varepsilon)\boldsymbol{p}_{0}+\varepsilon\boldsymbol{e}_{j}$, with $\boldsymbol{p}_{0}$ the null model or the Weibull model with $\eta=1.4$; running example, $n=200$, $4000$ replications.}
\smallskip
\small
\setlength{\tabcolsep}{4.5pt}
\begin{tabular}{ll rrrrrr r}
\toprule
 & & \multicolumn{6}{c}{$\beta$} & \\
\cmidrule{3-8}
Bulk, contaminated cell & Statistic & 0 & 0.2 & 0.4 & 0.6 & 0.8 & 1 & $\widehat{\beta}_{\mathrm{opt}}$ \\
\midrule
Null, first interval, $\varepsilon=0.05$ & $T^{\mathrm{CS}}$ & 0.235 & 0.207 & 0.190 & 0.179 & 0.182 & 0.189 & 0.185 \\
 & $T^{\mathrm{C}}_{1}$ & 0.138 & 0.124 & 0.123 & 0.130 & 0.142 & 0.162 & 0.125 \\
 & $X^{2}$ & 0.130 & 0.127 & 0.127 & 0.130 & 0.137 & 0.144 & 0.127 \\
\midrule
Null, first interval, $\varepsilon=0.10$ & $T^{\mathrm{CS}}$ & 0.700 & 0.651 & 0.613 & 0.591 & 0.590 & 0.607 & 0.624 \\
 & $T^{\mathrm{C}}_{1}$ & 0.518 & 0.464 & 0.443 & 0.455 & 0.496 & 0.546 & 0.439 \\
 & $X^{2}$ & 0.439 & 0.433 & 0.432 & 0.440 & 0.458 & 0.483 & 0.430 \\
\midrule
Null, third interval, $\varepsilon=0.05$ & $T^{\mathrm{CS}}$ & 0.342 & 0.324 & 0.295 & 0.278 & 0.267 & 0.261 & 0.338 \\
 & $T^{\mathrm{C}}_{1}$ & 0.212 & 0.185 & 0.168 & 0.163 & 0.166 & 0.167 & 0.199 \\
 & $X^{2}$ & 0.510 & 0.513 & 0.518 & 0.521 & 0.523 & 0.526 & 0.510 \\
\midrule
Null, third interval, $\varepsilon=0.10$ & $T^{\mathrm{CS}}$ & 0.882 & 0.871 & 0.855 & 0.836 & 0.821 & 0.814 & 0.878 \\
 & $T^{\mathrm{C}}_{1}$ & 0.784 & 0.740 & 0.698 & 0.669 & 0.657 & 0.653 & 0.765 \\
 & $X^{2}$ & 0.967 & 0.968 & 0.969 & 0.969 & 0.970 & 0.970 & 0.968 \\
\midrule
Null, survivors, $\varepsilon=0.05$ & $T^{\mathrm{CS}}$ & 0.044 & 0.047 & 0.047 & 0.048 & 0.048 & 0.049 & 0.043 \\
 & $T^{\mathrm{C}}_{1}$ & 0.049 & 0.050 & 0.049 & 0.049 & 0.049 & 0.050 & 0.049 \\
 & $X^{2}$ & 0.043 & 0.043 & 0.043 & 0.044 & 0.045 & 0.046 & 0.043 \\
\midrule
Weibull, first interval, $\varepsilon=0.05$ & $T^{\mathrm{CS}}$ & 0.261 & 0.278 & 0.290 & 0.299 & 0.306 & 0.314 & 0.286 \\
 & $T^{\mathrm{C}}_{1}$ & 0.241 & 0.266 & 0.279 & 0.291 & 0.296 & 0.304 & 0.282 \\
 & $X^{2}$ & 0.137 & 0.133 & 0.129 & 0.128 & 0.128 & 0.126 & 0.130 \\
\midrule
Weibull, third interval, $\varepsilon=0.05$ & $T^{\mathrm{CS}}$ & 0.816 & 0.780 & 0.734 & 0.695 & 0.658 & 0.635 & 0.816 \\
 & $T^{\mathrm{C}}_{1}$ & 0.538 & 0.412 & 0.305 & 0.246 & 0.211 & 0.197 & 0.537 \\
 & $X^{2}$ & 0.847 & 0.854 & 0.857 & 0.865 & 0.870 & 0.876 & 0.847 \\
\midrule
Weibull, survivors, $\varepsilon=0.05$ & $T^{\mathrm{CS}}$ & 0.243 & 0.212 & 0.184 & 0.157 & 0.141 & 0.131 & 0.244 \\
 & $T^{\mathrm{C}}_{1}$ & 0.177 & 0.150 & 0.130 & 0.118 & 0.108 & 0.100 & 0.179 \\
 & $X^{2}$ & 0.154 & 0.158 & 0.163 & 0.169 & 0.175 & 0.179 & 0.154 \\
\bottomrule
\end{tabular}
\end{table}

None of the tests has this stability, for any \(\beta\). With the bulk
at the null model, contamination of the first or the third interval
raises the rejection rates of all the statistics well above \(0.05\);
the cumulative statistics reject less than \(X^{2}\) when the third
interval is contaminated and more when the first is, so the comparison
of Section 4 depends on the position of the cell. Contamination of the
survivors is almost absorbed by the fit, since it lowers all the values
of the distribution function in nearly the same proportion: no test
detects it, and with Weibull data it reduces the power of every test.
Increasing \(\beta\) lowers the rejection rates of the cumulative
statistics under contamination of the null model only moderately, and
not monotonically: with \(\varepsilon=0.05\) in the first interval, the
rate of \(T^{\mathrm{CS}}\) falls from \(0.235\) for \(\beta=0\) to
\(0.179\) for \(\beta=0.6\) and rises again to \(0.189\) for
\(\beta=1\), and that of \(X^{2}\) hardly changes. With Weibull data and
contamination of the first interval, increasing \(\beta\) increases the
power of the cumulative statistics, from \(0.261\) to \(0.314\) for
\(T^{\mathrm{CS}}\), because the estimate is pulled less towards the
contaminated cell. The data-driven \(\widehat{\beta}_{\mathrm{opt}}\)
reacts to contamination of the first interval (mean \(0.22\) with
\(\varepsilon=0.05\) and \(0.34\) with \(\varepsilon=0.10\), against
\(0.08\) without contamination) but hardly to contamination of the third
interval or of the survivors. The estimator limits the influence of the
outlying counts on the fitted probabilities, but these counts still
enter the empirical distribution function, and hence the cumulative
residuals, without downweighting. Section 7 returns to this point.

The following result makes the last statement precise. Write the
statistic as \(T_{w,\beta}=n\tau(\widehat{\boldsymbol{p}})\), with the
functionals \[
\boldsymbol{r}(\boldsymbol{G})=\boldsymbol{C}\big(\boldsymbol{G}-\boldsymbol{\pi}\big(\boldsymbol{T}_{\beta}(\boldsymbol{G})\big)\big),\qquad
\tau(\boldsymbol{G})=\boldsymbol{r}^{\top}(\boldsymbol{G})\,\boldsymbol{W}\big(\boldsymbol{T}_{\beta}(\boldsymbol{G})\big)\,\boldsymbol{r}(\boldsymbol{G}),
\] defined for \(\boldsymbol{G}\in\mathcal{V}\cap\mathcal{P}_{L+1}\),
with \(\boldsymbol{T}_{\beta}\) and \(\mathcal{V}\) as in Proposition 1.

\textbf{Proposition 5.} \emph{Assume (A1)--(A4), let \(\beta\geq0\),
\(\boldsymbol{q}\in\mathcal{P}_{L+1}\), and
\(\boldsymbol{G}_{\varepsilon}=(1-\varepsilon)\boldsymbol{\pi}(\boldsymbol{\theta}^{\ast})+\varepsilon\boldsymbol{q}\),
\(\varepsilon\in[0,1]\). Then
\(g(\varepsilon)=\tau(\boldsymbol{G}_{\varepsilon})\) satisfies
\(g(0)=g'(0)=0\) and} \[
\mathrm{IF}_{2}(\boldsymbol{q};\tau,\beta)\overset{\Delta}{=}g''(0)=2\,\boldsymbol{h}_{\beta}^{\top}(\boldsymbol{q})\,\boldsymbol{W}(\boldsymbol{\theta}^{\ast})\,\boldsymbol{h}_{\beta}(\boldsymbol{q}),\qquad
\boldsymbol{h}_{\beta}(\boldsymbol{q})\overset{\Delta}{=}\boldsymbol{C}\boldsymbol{M}_{\beta}\big(\boldsymbol{q}-\boldsymbol{\pi}(\boldsymbol{\theta}^{\ast})\big).
\] \emph{For Pearson's statistic the same holds without
\(\boldsymbol{C}\) and with
\(\boldsymbol{D}^{-1}(\boldsymbol{\theta}^{\ast})\) in the place of
\(\boldsymbol{W}(\boldsymbol{\theta}^{\ast})\).}

\begin{remark}
The first-order influence function of the statistic vanishes, as for every quadratic statistic, and $\mathrm{IF}_{2}$ is the first term that measures the effect of contamination; with $\boldsymbol{q}=\boldsymbol{e}_{j}$ it is the effect of an outlying device in cell $j$. The tuning parameter enters only through $\boldsymbol{M}_{\beta}$:
$$
\boldsymbol{h}_{\beta}(\boldsymbol{e}_{j})=\boldsymbol{C}\big(\boldsymbol{e}_{j}-\boldsymbol{\pi}(\boldsymbol{\theta}^{\ast})\big)-\boldsymbol{C}\boldsymbol{J}(\boldsymbol{\theta}^{\ast})\boldsymbol{L}_{\beta}\big(\boldsymbol{e}_{j}-\boldsymbol{\pi}(\boldsymbol{\theta}^{\ast})\big).
$$
The first term, the cumulative residual of the outlying device itself, does not depend on $\beta$; the estimator changes only the second term, the correction due to the fit. A robust estimator therefore cannot make the test robust, which is what Table 5 shows. This is the difference from the Wald-type tests of hypotheses on the parameters: their first-order influence function also vanishes, but their $\mathrm{IF}_{2}$ is a quadratic form in the influence function of the estimator alone (Ghosh, Mandal, Martín and Pardo, 2016; Calviño, Martín and Pardo, 2021, Eq. (37)), so that a robust estimator yields a robust test, whereas for a goodness-of-fit statistic $\boldsymbol{h}_{\beta}(\boldsymbol{q})$ contains, in addition, the term $\boldsymbol{C}(\boldsymbol{q}-\boldsymbol{\pi}(\boldsymbol{\theta}^{\ast}))$, which does not involve the estimator. In the running example, $\mathrm{IF}_{2}$ of $T^{\mathrm{CS}}$ with $\beta=0$ is $9.65$, $12.85$ and $0.03$ for contamination of the first interval, the third interval and the survivors, and $\beta=0.4$ lowers it in some cells and raises it in others; the value for the survivors explains why their contamination is not detected. With contamination through the cells, the contaminating distribution ranges over the simplex $\mathcal{P}_{L+1}$, and the influence function of the estimator is bounded for every $\beta\geq0$, as noted by Balakrishnan, Jaenada and Pardo (2024, Section 7), who compare estimators through the gross-error sensitivity; the same holds for $\mathrm{IF}_{2}$ of the statistics. The comparison between $\beta=0$ and $\beta>0$ is therefore of the size of the influence functions, cell by cell (Lindsay, 1994), and in particular of their size when the probability of a cell, $\pi_{j}(\boldsymbol{\theta}^{\ast})$, is close to $0$, that is, when $\boldsymbol{\pi}(\boldsymbol{\theta}^{\ast})$ is close to the boundary of the simplex. For the estimator, $\beta>0$ replaces the factor $\pi_{j}^{-1}$ of $\boldsymbol{L}_{0}$ by $\pi_{j}^{\beta-1}$, which reduces the influence of a cell of small probability relative to the other cells. For a cell of moderate probability the comparison may go the other way: in the running example $\pi_{1}=0.082$ is one of the largest cell probabilities, and the influence of the first cell on the estimator is larger for $\beta=1$ than for $\beta=0$ (Figure 1a), whereas it is less than a third of it when $\pi_{1}$ is close to $0$ (Figure 2a). For the statistics the large factor comes from the weights, which do not depend on $\beta$: at $\boldsymbol{q}=\boldsymbol{e}_{j}$, $\mathrm{IF}_{2}$ of Pearson's statistic with $\beta=0$ is at most $2(1-\pi_{j})/\pi_{j}$, because $\boldsymbol{M}_{0}$ is an orthogonal projection in the inner product with matrix $\boldsymbol{D}^{-1}(\boldsymbol{\theta}^{\ast})$, and the weight $1/(F(1-F))$ of $T^{\mathrm{CS}}$ is large when $F_{\boldsymbol{\theta}^{\ast}}(t_{j})$ is close to $0$ or $1$. In the running example, moving the first inspection from $t_{1}=10$ to $t_{1}=0.1$, so that $\pi_{1}$ falls from $0.082$ to $0.0009$, raises $\mathrm{IF}_{2}(\boldsymbol{e}_{1})$ from $9.65$ to about $2300$ for $T^{\mathrm{CS}}$ and from $12.8$ to about $2300$ for $X^{2}$, for each $\beta\in\{0,0.4,1\}$, whereas for $T^{\mathrm{C}}_{1}$, whose weight is constant, it stays below $9$. Tests whose level is stable under contamination need a statistic that downweights the first term.
\end{remark}

Figure 1 shows these influence functions cell by cell in the running
example, with the norm of the influence function of the estimator
computed in the log-rates. The tuning parameter changes the influence
function of the estimator markedly (panel a), but it changes
\(\mathrm{IF}_{2}\) of the cumulative statistics only moderately, and in
both directions depending on the cell (panels b and c); for \(X^{2}\)
the curves coincide (panel d). In Figure 2,
\(\boldsymbol{\theta}^{\ast}\) is kept fixed and the first inspection
time is moved towards the origin, so that
\(\pi_{1}(\boldsymbol{\theta}^{\ast})\) tends to \(0\). The influence of
the first cell on the estimator is smaller for \(\beta=1\) than for
\(\beta=0\) once \(\pi_{1}\) is below about \(0.05\), and less than a
third of it as \(\pi_{1}\) tends to \(0\), whereas at the design value
\(\pi_{1}=0.082\) it is larger; whereas \(\mathrm{IF}_{2}\) of
\(T^{\mathrm{CS}}\) and \(X^{2}\) grows like \(\pi_{1}^{-1}\) for both
values of \(\beta\), and that of \(T^{\mathrm{C}}_{1}\) stays bounded.

\begin{figure}[htbp]
\centering
\includegraphics[width=\textwidth]{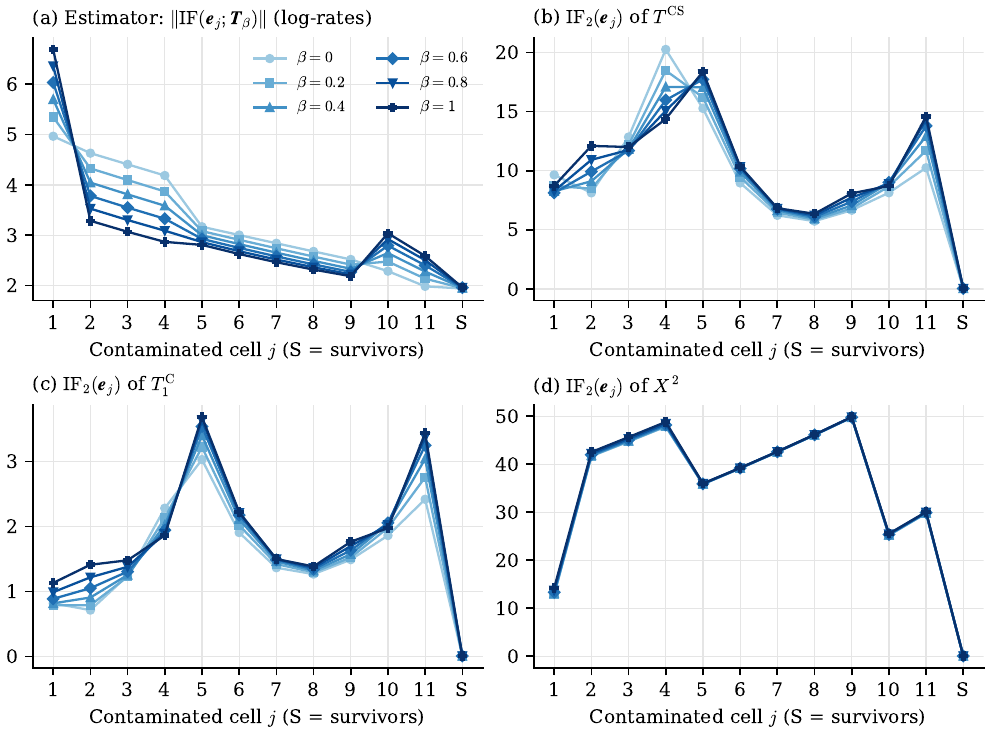}
\caption{Influence functions at $\boldsymbol{q}=\boldsymbol{e}_{j}$, cell by cell, for $\beta\in\{0,0.2,0.4,0.6,0.8,1\}$, running example of Section 3.4: (a) norm of the influence function of the estimator, in the log-rates; (b)--(d) second-order influence function $\mathrm{IF}_{2}$ of Proposition 5 for $T^{\mathrm{CS}}$, $T^{\mathrm{C}}_{1}$ and $X^{2}$.}
\end{figure}

\begin{figure}[htbp]
\centering
\includegraphics[width=\textwidth]{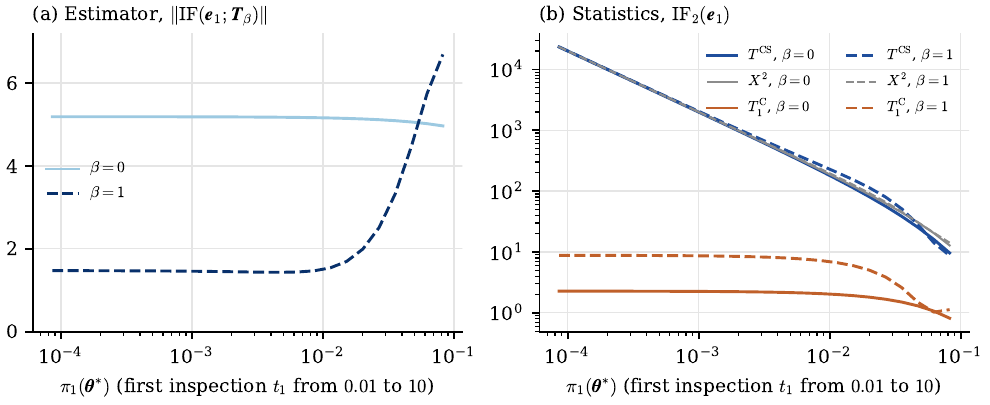}
\caption{Influence of a device in the first cell as $\pi_{1}(\boldsymbol{\theta}^{\ast})$ tends to $0$, with $\boldsymbol{\theta}^{\ast}$ fixed and the first inspection time $t_{1}$ moved from $10$ to $0.01$ in the running example, for $\beta=0$ (solid) and $\beta=1$ (dashed): (a) the estimator; (b) $\mathrm{IF}_{2}$ of the statistics, on a logarithmic scale.}
\end{figure}

The minimum density power divergence estimator is nevertheless useful
for the test. Under contamination of fixed size, the observed proportion
of the contaminated cell is far from its probability under the model,
and the estimating equations with \(\beta>0\) give it less weight; this
is a non-local effect, not described by the influence function. A fit
with \(\beta>0\) is then pulled less towards a contaminated cell that
drives the estimate, so the contamination stays in the residuals instead
of being absorbed by the parameters, and it does not mask a departure
from the model. With Weibull data and contamination of the first
interval, the power of \(T^{\mathrm{CS}}\) increases from \(0.261\) for
\(\beta=0\) to \(0.314\) for \(\beta=1\), and that of
\(T^{\mathrm{C}}_{1}\) from \(0.241\) to \(0.304\) (Table 5), whereas
\(X^{2}\) does not benefit; the cumulative residuals relative to the
robust fit also point to the contaminated cell. The data-driven
\(\widehat{\beta}_{\mathrm{opt}}\) selects \(\beta>0\) precisely in this
situation. Moreover, an analysis that estimates the parameters with
\(\beta>0\), to extrapolate robustly to the normal stress, needs a
goodness-of-fit test calibrated for that estimator: with the critical
values for \(\beta=0\), the asymptotic sizes of \(T^{\mathrm{CS}}\) and
\(T^{\mathrm{C}}_{1}\) after a fit with \(\beta=1\) would be \(0.073\)
and \(0.101\) (Section 3.4).

\hypertarget{application-light-bulbs-under-step-voltage}{%
\section{Application: light bulbs under step
voltage}\label{application-light-bulbs-under-step-voltage}}

Zhu (2010, Table 5.4), as reproduced by Jaenada, Millán and Pardo
(2026), reports a simple step-stress test of \(n=64\) miniature light
bulbs at \(2.25\) V on \([0,96]\) h and \(2.44\) V on \((96,140]\) h,
designed for \(2\) V; 53 failure times are recorded and 11 bulbs
survive. The failure mode studied is the long-term failure of the
filament. The failures were recorded continuously, so the inspection
grids below are ours, chosen to illustrate interval monitoring. In Table
5.4 of Zhu (2010) the failure at \(102.10\) h is coded as a failure at
the first stress level; since it occurs after \(\tau_{1}=96\), we assign
it to the second level, as Jaenada, Millán and Pardo (2026) do. Zhu
(2010, Section 5.4) fits the exponential cumulative exposure model with
a log-linear link in the voltage, that is, the null model of this paper,
and uses it to extrapolate the mean lifetime at \(2\) V, without
checking the fit. We group the failure times at three inspection grids,
each containing \(\tau_{1}=96\) and \(\tau_{2}=140\).

Grid A has inspections every 12 h up to 96 h and then at 107, 118, 129
and 140 h (\(L=12\)), with counts \(0,8,6,7,3,2,3,5,5,4,7,3,11\); grid B
has inspections every 16 h up to 96 h and then at 107, 118, 129 and 140
h (\(L=10\)), with counts \(2,11,8,4,3,6,5,4,7,3,11\); and grid C has
inspections at 24, 48, 72, 96, 118 and 140 h (\(L=6\)), with counts
\(8,13,5,8,9,10,11\).

For voltage the inverse power relationship, log-linear in the logarithm
of the voltage, is often preferred to the log-linear link (Escobar and
Meeker, 2006, Section 8.2). With two stress levels the choice does not
matter: the inverse power link
\(\lambda_{i}=\theta_{0}x_{i}^{\theta_{1}}\), like the log-linear one,
maps its parameter space one-to-one onto \((0,\infty)^{2}\) (Section
2.2), so the null hypothesis and the tests are the same. With grid B the
maximum likelihood estimates of the mean lifetimes are \(130.7\) h and
\(46.4\) h, against \(131.6\) h and \(46.5\) h from the exact times
(Jaenada, Millán and Pardo, 2026). Table 6 gives the \(p\)-values of
Corollary 1; with \(\beta=0\) the statistics are
\(T^{\mathrm{CS}}=7.912\), \(T^{\mathrm{C}}_{1}=0.803\) and
\(X^{2}=14.566\) with grid A.

\begin{table}[htbp]
\centering
\caption{Light-bulb data: asymptotic $p$-values of Corollary 1 for the exponential cumulative exposure model, for each $\beta$ of the grid and for the data-driven $\widehat{\beta}_{\mathrm{opt}}$, which equals $0$ for the three grids.}
\smallskip
\small
\begin{tabular}{ll rrrrrr r}
\toprule
 & & \multicolumn{6}{c}{$\beta$} & \\
\cmidrule{3-8}
Grid & Statistic & 0 & 0.2 & 0.4 & 0.6 & 0.8 & 1 & $\widehat{\beta}_{\mathrm{opt}}$ \\
\midrule
A ($L=12$) & $T^{\mathrm{CS}}$ & 0.080 & 0.077 & 0.077 & 0.078 & 0.081 & 0.085 & 0.080 \\
 & $T^{\mathrm{C}}_{1}$ & 0.261 & 0.264 & 0.272 & 0.283 & 0.295 & 0.307 & 0.261 \\
 & $X^{2}$ & 0.149 & 0.150 & 0.151 & 0.153 & 0.155 & 0.157 & 0.149 \\
\midrule
B ($L=10$) & $T^{\mathrm{CS}}$ & 0.117 & 0.116 & 0.117 & 0.120 & 0.125 & 0.132 & 0.117 \\
 & $T^{\mathrm{C}}_{1}$ & 0.282 & 0.288 & 0.298 & 0.311 & 0.326 & 0.341 & 0.282 \\
 & $X^{2}$ & 0.133 & 0.133 & 0.134 & 0.135 & 0.137 & 0.138 & 0.133 \\
\midrule
C ($L=6$) & $T^{\mathrm{CS}}$ & 0.410 & 0.400 & 0.392 & 0.386 & 0.382 & 0.379 & 0.410 \\
 & $T^{\mathrm{C}}_{1}$ & 0.469 & 0.462 & 0.456 & 0.452 & 0.449 & 0.447 & 0.469 \\
 & $X^{2}$ & 0.232 & 0.232 & 0.233 & 0.234 & 0.234 & 0.235 & 0.232 \\
\bottomrule
\end{tabular}
\end{table}

No test rejects at level \(0.05\) for any \(\beta\), the \(p\)-values
change little over the grid, and the data-driven criterion selects the
maximum likelihood estimator for the three grids, so no count has a
large influence on the fit. The finest grid shows the most evidence
against the model, and \(T^{\mathrm{CS}}\) more than \(X^{2}\)
(\(p=0.08\) against \(0.15\)). The cumulative residuals locate the
discrepancy at the start of the test: no failure in \((0,12]\), against
\(5.6\) expected, and \(8\) failures in \((12,24]\), against \(5.1\).
Coarser grids merge these cells and lose this evidence.

Two explanations fit this pattern. One is a failure rate that increases
with age: the Weibull cumulative exposure model gives
\(\widehat{\eta}=1.395\) from the exact times, with likelihood ratio
test of \(\eta=1\) giving \(p=0.038\), and \(\widehat{\eta}=1.396\)
(\(p=0.040\)) with grid A; the likelihood ratio test is directed at this
alternative and, among the omnibus tests, \(T^{\mathrm{CS}}\) comes
closest, in agreement with Section 4. The other is an anomaly of the
early cells of this sample. Zhu (2010, Table 5.3) also reports
constant-voltage tests on bulbs of the same type, \(47\) bulbs at
\(2.20\) V censored at \(160.13\) h (\(25\) failures) and \(22\) bulbs
at \(2.46\) V censored at \(130.47\) h (\(21\) failures). In both, the
first failure occurs before \(4\) h, and the Weibull fit gives no
evidence against exponential lifetimes (\(\widehat{\eta}=0.94\),
\(p=0.71\), and \(\widehat{\eta}=1.15\), \(p=0.46\), by the likelihood
ratio test). Interpolated on the log-linear link, their rates give mean
lifetimes of \(152\) h at \(2.25\) V and \(57\) h at \(2.44\) V, against
\(131\) h and \(46\) h from the step-stress data, which is compatible
with these sample sizes. The step-stress data alone cannot separate the
two explanations; the cumulative statistics point to where the model
fails, which is what a goodness-of-fit check should deliver before the
model is used for extrapolation to \(2\) V.

\hypertarget{conclusions}{%
\section{Conclusions}\label{conclusions}}

Cumulative residuals are a natural basis for goodness-of-fit testing in
interval-monitored step-stress tests: they respect the order of the
cells, and they are more powerful than Pearson's statistic against a
wrong lifetime distribution. With parameters estimated by the minimum
density power divergence estimator their limit is a weighted chi-square
distribution with \(L-s\) terms, and plugging in the estimated weights
gives a test with asymptotic size \(\alpha\). In the absence of
contamination, the gain over Pearson's statistic is largest when
\(\beta\) is close to \(0\); in a design with similar rates at the two
stress levels it disappears for \(\beta\geq0.6\), whereas in the
light-bulb design it does not depend on \(\beta\), and the data-driven
choice of Basak, Basu and Jones (2021) selects small values of \(\beta\)
unless an early cell is contaminated. When contamination is suspected,
\(\beta>0\) is recommended: it reduces the influence of the early cells
on the fit, it prevents the contamination from masking a departure from
the model, and the calibration of Corollary 1 accounts for the value of
\(\beta\). The theory holds for any smooth parametric model that
satisfies (A1)--(A4), and so extends directly to Weibull lifetimes and
to multiple step-stress designs, where the link with the stress becomes
testable. The minimum density power divergence estimator makes the fit
robust, but not the test: the simulation shows that the size of every
statistic considered here is inflated when a fixed proportion of the
devices fails in an early interval, whatever \(\beta\). Robust
goodness-of-fit tests require a statistic that also downweights the
outlying cells; this is the subject of ongoing work. Other extensions
are the choice of the inspection grid, goodness-of-fit tests with exact
failure times, and joint analyses of constant-stress and step-stress
tests on the same units, such as those of Zhu (2010) at four voltages,
in which the link with the stress and the cumulative exposure rule
become testable.

\hypertarget{references}{%
\section*{References}\label{references}}
\addcontentsline{toc}{section}{References}

\begingroup\small\setlength{\parskip}{3pt}

\noindent\hangindent=1.5em\hangafter=1Balakrishnan, N., Jaenada, M. and
Pardo, L. (2024). Robust inference for an interval-monitored step-stress
experiment with competing risks for failure with an application to
capacitor data. \emph{Computers \& Industrial Engineering}, 197, 110536.

\noindent\hangindent=1.5em\hangafter=1Balakrishnan, N., Jaenada, M. and
Pardo, L. (2026a). Robust inference for step-stress experiments under
interval-censoring. \emph{TEST}. doi:10.1007/s11749-026-01009-8.

\noindent\hangindent=1.5em\hangafter=1Balakrishnan, N., Jaenada, M. and
Pardo, L. (2026b). Robust inference for intermittently-monitored
step-stress tests under Weibull lifetime distributions. \emph{Quality
and Reliability Engineering International}, 42(2), 837--857.
doi:10.1002/qre.70113.

\noindent\hangindent=1.5em\hangafter=1Basak, S., Basu, A. and Jones, M.
C. (2021). On the `optimal' density power divergence tuning parameter.
\emph{Journal of Applied Statistics}, 48(3), 536--556.

\noindent\hangindent=1.5em\hangafter=1Basu, A., Harris, I. R., Hjort, N.
L. and Jones, M. C. (1998). Robust and efficient estimation by
minimising a density power divergence. \emph{Biometrika}, 85(3),
549--559.

\noindent\hangindent=1.5em\hangafter=1Basu, A., Mandal, A., Martín, N.
and Pardo, L. (2018). Testing composite hypothesis based on the density
power divergence. \emph{Sankhya B}, 80(2), 222--262.

\noindent\hangindent=1.5em\hangafter=1Birch, M. W. (1964). A new proof
of the Pearson--Fisher theorem. \emph{The Annals of Mathematical
Statistics}, 35(2), 817--824.

\noindent\hangindent=1.5em\hangafter=1Calviño, A., Martín, N. and Pardo,
L. (2021). Robustness of minimum density power divergence estimators and
Wald-type test statistics in loglinear models with multinomial sampling.
\emph{Journal of Computational and Applied Mathematics}, 386, 113214.

\noindent\hangindent=1.5em\hangafter=1Choulakian, V., Lockhart, R. A.
and Stephens, M. A. (1994). Cramér--von Mises statistics for discrete
distributions. \emph{The Canadian Journal of Statistics}, 22(1),
125--137.

\noindent\hangindent=1.5em\hangafter=1Escobar, L. A. and Meeker, W. Q.
(2006). A review of accelerated test models. \emph{Statistical Science},
21(4), 552--577.

\noindent\hangindent=1.5em\hangafter=1Ghosh, A., Mandal, A., Martín, N.
and Pardo, L. (2016). Influence analysis of robust Wald-type tests.
\emph{Journal of Multivariate Analysis}, 147, 102--126.

\noindent\hangindent=1.5em\hangafter=1Hamada, M. and Wu, C. F. J.
(1990). A critical look at accumulation analysis and related methods
(with discussion). \emph{Technometrics}, 32(2), 119--162.

\noindent\hangindent=1.5em\hangafter=1Hirotsu, C. (1986). Cumulative
chi-squared statistic as a tool for testing goodness of fit.
\emph{Biometrika}, 73(1), 165--173.

\noindent\hangindent=1.5em\hangafter=1Imhof, J. P. (1961). Computing the
distribution of quadratic forms in normal variables. \emph{Biometrika},
48(3/4), 419--426.

\noindent\hangindent=1.5em\hangafter=1Jaenada, M. (2024). \emph{Robust
Inference for Step-Stress Accelerated Life Tests under Interval
Censoring}. Doctoral thesis, Complutense University of Madrid.

\noindent\hangindent=1.5em\hangafter=1Jaenada, M., Millán, J. M. and
Pardo, L. (2026). Robust tests for step-stress models under exponential
lifetimes. arXiv:2609.17827.

\noindent\hangindent=1.5em\hangafter=1Kundu, D. and Ganguly, A. (2017).
\emph{Analysis of Step-Stress Models: Existing Results and Some Recent
Developments}. Academic Press, London.

\noindent\hangindent=1.5em\hangafter=1Lindsay, B. G. (1994). Efficiency
versus robustness: the case for minimum Hellinger distance and related
methods. \emph{The Annals of Statistics}, 22(2), 1081--1114.

\noindent\hangindent=1.5em\hangafter=1Lockhart, R. A., Spinelli, J. J.
and Stephens, M. A. (2007). Cramér--von Mises statistics for discrete
distributions with unknown parameters. \emph{The Canadian Journal of
Statistics}, 35(1), 125--133.

\noindent\hangindent=1.5em\hangafter=1Martín, N. and de Frutos, J. M.
(2026). Cumulative \(\phi\)-divergences for ordinal multinomial data,
with an application to rank histograms. Invited chapter in
\emph{Advances in Statistical Modeling and Applications}, Springer.

\noindent\hangindent=1.5em\hangafter=1Nair, V. N. (1986). Testing in
industrial experiments with ordered categorical data.
\emph{Technometrics}, 28(4), 283--291.

\noindent\hangindent=1.5em\hangafter=1Nair, V. N. (1987).
Chi-squared-type tests for ordered alternatives in contingency tables.
\emph{Journal of the American Statistical Association}, 82(397),
283--291.

\noindent\hangindent=1.5em\hangafter=1Nelson, W. (1980). Accelerated
life testing: step-stress models and data analyses. \emph{IEEE
Transactions on Reliability}, R-29(2), 103--108.

\noindent\hangindent=1.5em\hangafter=1Nelson, W. (1990).
\emph{Accelerated Testing: Statistical Models, Test Plans, and Data
Analyses}. Wiley, New York.

\noindent\hangindent=1.5em\hangafter=1Sedyakin, N. M. (1966). On one
physical principle in reliability theory. \emph{Technical Cybernetics},
3, 80--87.

\noindent\hangindent=1.5em\hangafter=1Spinelli, J. J. (2001). Testing
fit for the grouped exponential distribution. \emph{The Canadian Journal
of Statistics}, 29(3), 451--458.

\noindent\hangindent=1.5em\hangafter=1Taguchi, G. (1974). A new
statistical analysis for clinical data, the accumulating analysis, in
contrast with the chi-square test. \emph{Saishin Igaku}, 29, 806--813.

\noindent\hangindent=1.5em\hangafter=1Warwick, J. and Jones, M. C.
(2005). Choosing a robustness tuning parameter. \emph{Journal of
Statistical Computation and Simulation}, 75(7), 581--588.

\noindent\hangindent=1.5em\hangafter=1Zhu, Y. (2010). \emph{Optimal
Design and Equivalency of Accelerated Life Testing Plans}. Doctoral
thesis, Rutgers, The State University of New Jersey.

\noindent\hangindent=1.5em\hangafter=1Zografos, K. (2023). On
reconsidering entropies and divergences and their cumulative
counterparts: Csiszár's, DPD's and Fisher's type cumulative and survival
measures. \emph{Probability in the Engineering and Informational
Sciences}, 37(1), 294--321.

\endgroup

\hypertarget{appendix-proofs}{%
\section*{Appendix: proofs}\label{appendix-proofs}}
\addcontentsline{toc}{section}{Appendix: proofs}

\emph{Proof of Lemma 1.} (A1) and (A2): \(\Theta\) is open, and
\(u(t;\boldsymbol{\theta})\) is strictly increasing in \(t\) because
\(\lambda_{i}(\boldsymbol{\theta})>0\), so
\(\pi_{j}(\boldsymbol{\theta})>0\) for \(j\leq L\) and
\(\pi_{L+1}(\boldsymbol{\theta})=e^{-u(t_{L};\boldsymbol{\theta})}>0\).
(A3): \(\boldsymbol{\pi}(\boldsymbol{\theta})\) is infinitely
differentiable. The columns of \(\boldsymbol{J}(\boldsymbol{\theta})\)
lie in \(\mathcal{S}_{0}\), on which \(\boldsymbol{C}\) is injective
(proof of Lemma 2), so
\(\operatorname{rank}\boldsymbol{J}(\boldsymbol{\theta})=\operatorname{rank}\boldsymbol{C}\boldsymbol{J}(\boldsymbol{\theta})\);
row \(j\) of \(\boldsymbol{C}\boldsymbol{J}(\boldsymbol{\theta})\) is
\(e^{-u_{j}}\partial u_{j}/\partial\boldsymbol{\theta}^{\top}\),
proportional to \((1,\theta_{0}\bar{x}_{j})\) with
\(\bar{x}_{j}=\sum_{l}x_{l}\lambda_{l}\Delta_{l}(t_{j})/\sum_{l}\lambda_{l}\Delta_{l}(t_{j})\)
and \(\Delta_{l}(t)\) the time spent at \(x_{l}\) by \(t\). Since
\(\bar{x}_{j}=x_{1}\) for \(j\leq m\) and \(\bar{x}_{j}>x_{1}\) for
\(j>m\), the rank is \(2\). (A4):
\(\boldsymbol{\theta}\mapsto(\lambda_{1},\lambda_{2})\) is a
homeomorphism of \(\Theta\) onto \((0,\infty)^{2}\), with inverse
\(\theta_{1}=\log(\lambda_{2}/\lambda_{1})/(x_{2}-x_{1})\),
\(\theta_{0}=\lambda_{1}e^{-\theta_{1}x_{1}}\). Let
\((\boldsymbol{\theta}_{r})_{r\in\mathbb{N}}\) be a sequence in
\(\Theta\) with
\(\boldsymbol{\pi}(\boldsymbol{\theta}_{r})\xrightarrow[r\to\infty]{}\boldsymbol{\pi}(\boldsymbol{\theta}^{\ast})\).
If a subsequence of \((\lambda_{1r},\lambda_{2r})\) converges in
\((0,\infty)^{2}\), its limit has cell probabilities
\(\boldsymbol{\pi}(\boldsymbol{\theta}^{\ast})\), and since
\(F(t_{1})=1-e^{-\lambda_{1}t_{1}}\) determines \(\lambda_{1}\) and then
\(F(t_{L})\) determines \(\lambda_{2}\), the limit is
\((\lambda_{1}(\boldsymbol{\theta}^{\ast}),\lambda_{2}(\boldsymbol{\theta}^{\ast}))\).
Otherwise a subsequence leaves every compact subset of
\((0,\infty)^{2}\), and along a further subsequence either
\(\lambda_{1r}\xrightarrow[r\to\infty]{}0\)
(\(\pi_{1}(\boldsymbol{\theta}_{r})\xrightarrow[r\to\infty]{}0\)), or
\(\lambda_{1r}\xrightarrow[r\to\infty]{}\infty\)
(\(\pi_{1}(\boldsymbol{\theta}_{r})\xrightarrow[r\to\infty]{}1\)), or
\(\lambda_{1r}\) converges in \((0,\infty)\) and
\(\lambda_{2r}\xrightarrow[r\to\infty]{}0\)
(\(\pi_{m+1}(\boldsymbol{\theta}_{r})\xrightarrow[r\to\infty]{}0\)) or
\(\lambda_{2r}\xrightarrow[r\to\infty]{}\infty\)
(\(\pi_{L+1}(\boldsymbol{\theta}_{r})\xrightarrow[r\to\infty]{}0\)),
each impossible because \(0<\pi_{j}(\boldsymbol{\theta}^{\ast})<1\).
Hence
\(\boldsymbol{\theta}_{r}\xrightarrow[r\to\infty]{}\boldsymbol{\theta}^{\ast}\).
\(\square\)

\emph{Proof of Proposition 1.} For \(\beta=0\) this is Birch's theorem
(Birch, 1964), whose conditions are (A1)--(A4); we give the proof for
\(\beta>0\).

We first bound the effect of a change of the first argument. Since
\(0\leq\pi_{j}(\boldsymbol{\theta})\leq1\) and
\(|p_{j}^{1+\beta}-q_{j}^{1+\beta}|\leq(1+\beta)|p_{j}-q_{j}|\), for
\(\boldsymbol{p},\boldsymbol{q}\in\mathcal{P}_{L+1}\) and all
\(\boldsymbol{\theta}\), \[
\big|d_{\beta}(\boldsymbol{p},\boldsymbol{\pi}(\boldsymbol{\theta}))-d_{\beta}(\boldsymbol{q},\boldsymbol{\pi}(\boldsymbol{\theta}))\big|\leq c_{\beta}\|\boldsymbol{p}-\boldsymbol{q}\|_{1},\qquad c_{\beta}=2(1+\beta)/\beta.
\tag{A.1}
\] Next, the true model is separated from the models outside any
neighbourhood of \(\boldsymbol{\theta}^{\ast}\): for every open
neighbourhood \(\mathcal{U}'\) of \(\boldsymbol{\theta}^{\ast}\), \[
\delta(\mathcal{U}')\overset{\Delta}{=}\inf_{\boldsymbol{\theta}\in\Theta\setminus\mathcal{U}'}d_{\beta}\big(\boldsymbol{\pi}(\boldsymbol{\theta}^{\ast}),\boldsymbol{\pi}(\boldsymbol{\theta})\big)>0.
\tag{A.2}
\] Otherwise there are \(\boldsymbol{\theta}_{r}\notin\mathcal{U}'\)
with
\(d_{\beta}(\boldsymbol{\pi}(\boldsymbol{\theta}^{\ast}),\boldsymbol{\pi}(\boldsymbol{\theta}_{r}))\xrightarrow[r\to\infty]{}0\);
by compactness of \(\mathcal{P}_{L+1}\) a subsequence of
\(\boldsymbol{\pi}(\boldsymbol{\theta}_{r})\) converges to some
\(\boldsymbol{q}\) with
\(d_{\beta}(\boldsymbol{\pi}(\boldsymbol{\theta}^{\ast}),\boldsymbol{q})=0\),
so \(\boldsymbol{q}=\boldsymbol{\pi}(\boldsymbol{\theta}^{\ast})\) (Basu
et al., 1998), and by (A4) that subsequence of
\(\boldsymbol{\theta}_{r}\) converges to
\(\boldsymbol{\theta}^{\ast}\in\mathcal{U}'\), a contradiction.

We now locate the critical points near \(\boldsymbol{\theta}^{\ast}\).
By (A2)--(A3), \(\boldsymbol{\Psi}_{\beta}\) in (3) is continuously
differentiable on \(\Theta\times\mathbb{R}^{L+1}\),
\(\boldsymbol{\Psi}_{\beta}(\boldsymbol{\theta}^{\ast},\boldsymbol{\pi}(\boldsymbol{\theta}^{\ast}))=\boldsymbol{0}\),
\(\partial\boldsymbol{\Psi}_{\beta}/\partial\boldsymbol{\theta}^{\top}=-\boldsymbol{A}_{\beta}\)
at
\((\boldsymbol{\theta}^{\ast},\boldsymbol{\pi}(\boldsymbol{\theta}^{\ast}))\),
because the terms with derivatives of
\(\boldsymbol{J}^{\top}(\boldsymbol{\theta})\boldsymbol{D}^{\beta-1}(\boldsymbol{\theta})\)
are multiplied by
\(\boldsymbol{p}-\boldsymbol{\pi}(\boldsymbol{\theta})\), and
\(\partial\boldsymbol{\Psi}_{\beta}/\partial\boldsymbol{p}^{\top}=\boldsymbol{J}^{\top}(\boldsymbol{\theta})\boldsymbol{D}^{\beta-1}(\boldsymbol{\theta})\).
\(\boldsymbol{A}_{\beta}\) is positive definite by (A2)--(A3). By the
implicit function theorem there are open neighbourhoods
\(\mathcal{U}_{0}\) of \(\boldsymbol{\theta}^{\ast}\), with compact
closure in \(\Theta\), and \(\mathcal{V}_{0}\) of
\(\boldsymbol{\pi}(\boldsymbol{\theta}^{\ast})\), and a continuously
differentiable
\(\boldsymbol{T}_{\beta}:\mathcal{V}_{0}\to\mathcal{U}_{0}\) such that,
on \(\mathcal{U}_{0}\times\mathcal{V}_{0}\),
\(\boldsymbol{\Psi}_{\beta}(\boldsymbol{\theta},\boldsymbol{p})=\boldsymbol{0}\)
if and only if
\(\boldsymbol{\theta}=\boldsymbol{T}_{\beta}(\boldsymbol{p})\).
Differentiating
\(\boldsymbol{\Psi}_{\beta}(\boldsymbol{T}_{\beta}(\boldsymbol{p}),\boldsymbol{p})=\boldsymbol{0}\)
at \(\boldsymbol{\pi}(\boldsymbol{\theta}^{\ast})\) gives
\(\partial\boldsymbol{T}_{\beta}/\partial\boldsymbol{p}^{\top}=\boldsymbol{L}_{\beta}\).

Finally, choose an open \(\mathcal{U}\ni\boldsymbol{\theta}^{\ast}\)
with \(\overline{\mathcal{U}}\subset\mathcal{U}_{0}\), let
\(\delta=\delta(\mathcal{U})\) be given by (A.2), and
\(\mathcal{V}=\{\boldsymbol{p}\in\mathcal{V}_{0}:c_{\beta}\|\boldsymbol{p}-\boldsymbol{\pi}(\boldsymbol{\theta}^{\ast})\|_{1}<\delta/3\}\).
For \(\boldsymbol{p}\in\mathcal{V}\cap\mathcal{P}_{L+1}\), (A.1) gives
\(d_{\beta}(\boldsymbol{p},\boldsymbol{\pi}(\boldsymbol{\theta}^{\ast}))<\delta/3\)
and
\(d_{\beta}(\boldsymbol{p},\boldsymbol{\pi}(\boldsymbol{\theta}))>2\delta/3\)
for \(\boldsymbol{\theta}\notin\mathcal{U}\). The minimum of
\(d_{\beta}(\boldsymbol{p},\boldsymbol{\pi}(\boldsymbol{\theta}))\) over
the compact \(\overline{\mathcal{U}}\) is attained, at a point of
\(\mathcal{U}\), and is the global minimum over \(\Theta\); every global
minimiser lies in the open set \(\mathcal{U}\), is a critical point, and
so equals \(\boldsymbol{T}_{\beta}(\boldsymbol{p})\). \(\square\)

\emph{Proof of Proposition 2.} Since
\(\boldsymbol{p}_{n}\xrightarrow[n\to\infty]{}\boldsymbol{\pi}(\boldsymbol{\theta}^{\ast})\)
and
\(\operatorname{E}\big[\|\widehat{\boldsymbol{p}}-\boldsymbol{p}_{n}\|^{2}\big]\leq1/n\),
\(\widehat{\boldsymbol{p}}\xrightarrow[n\to\infty]{P}\boldsymbol{\pi}(\boldsymbol{\theta}^{\ast})\),
so
\(\operatorname{P}(\widehat{\boldsymbol{p}}\in\mathcal{V})\xrightarrow[n\to\infty]{}1\),
and on that event
\(\widehat{\boldsymbol{\theta}}_{\beta}=\boldsymbol{T}_{\beta}(\widehat{\boldsymbol{p}})\)
is the unique minimiser by Proposition 1. By the differentiability of
\(\boldsymbol{T}_{\beta}\) at
\(\boldsymbol{\pi}(\boldsymbol{\theta}^{\ast})\), \[
\boldsymbol{T}_{\beta}(\widehat{\boldsymbol{p}})-\boldsymbol{\theta}^{\ast}=\boldsymbol{L}_{\beta}\big(\widehat{\boldsymbol{p}}-\boldsymbol{\pi}(\boldsymbol{\theta}^{\ast})\big)+R\big(\|\widehat{\boldsymbol{p}}-\boldsymbol{\pi}(\boldsymbol{\theta}^{\ast})\|\big)\big\|\widehat{\boldsymbol{p}}-\boldsymbol{\pi}(\boldsymbol{\theta}^{\ast})\big\|,
\] with
\(R(\|\boldsymbol{p}-\boldsymbol{\pi}(\boldsymbol{\theta}^{\ast})\|)\to\boldsymbol{0}\)
as \(\boldsymbol{p}\to\boldsymbol{\pi}(\boldsymbol{\theta}^{\ast})\).
Multiplying by \(\sqrt{n}\), the remainder tends to \(\boldsymbol{0}\)
in probability because
\(\sqrt{n}\|\widehat{\boldsymbol{p}}-\boldsymbol{\pi}(\boldsymbol{\theta}^{\ast})\|\)
is bounded in probability and
\(R(\|\widehat{\boldsymbol{p}}-\boldsymbol{\pi}(\boldsymbol{\theta}^{\ast})\|)\xrightarrow[n\to\infty]{P}\boldsymbol{0}\).
Under \(H_{0}\) the strong law of large numbers gives
\(\widehat{\boldsymbol{p}}\in\mathcal{V}\) for all \(n\) large enough
almost surely, hence almost sure consistency, and the central limit
theorem
\(\sqrt{n}(\widehat{\boldsymbol{p}}-\boldsymbol{\pi}(\boldsymbol{\theta}^{\ast}))\xrightarrow[n\to\infty]{\mathcal{L}}\mathcal{N}_{L+1}(\boldsymbol{0},\boldsymbol{\Sigma}(\boldsymbol{\theta}^{\ast}))\)
gives the limit, with
\(\boldsymbol{L}_{\beta}\boldsymbol{\Sigma}(\boldsymbol{\theta}^{\ast})\boldsymbol{L}_{\beta}^{\top}=\boldsymbol{A}_{\beta}^{-1}\boldsymbol{K}_{\beta}\boldsymbol{A}_{\beta}^{-1}\).
\(\square\)

\textbf{Proposition 3 (existence of the maximum likelihood estimator).}
\emph{Let \(S=n-\sum_{j\leq m}N_{j}\). The likelihood factorises as
\(L_{1}(\lambda_{1})L_{2}(\lambda_{2})\), where \(L_{1}\) is the
multinomial likelihood of \((N_{1},\ldots,N_{m},S)\) and \(L_{2}\) the
multinomial likelihood of \((N_{m+1},\ldots,N_{L+1})\) given \(S\). The
maximum likelihood estimator exists in \(\Theta\), and is unique, if and
only if \(N_{1}+\cdots+N_{m}\geq1\), \(N_{1}<n\),
\(N_{m+1}+\cdots+N_{L}\geq1\) and \(N_{m+1}<S\).}

\emph{Proof.} For \(j\leq m\) the cell probabilities depend only on
\(\lambda_{1}\), and so does
\(\operatorname{P}(T>\tau_{1})=e^{-\lambda_{1}\tau_{1}}\); for \(j>m\),
\(\pi_{j}=e^{-\lambda_{1}\tau_{1}}\operatorname{P}(t_{j-1}<T\leq t_{j}\mid T>\tau_{1})\),
and by the lack of memory of the exponential distribution the
conditional probabilities depend only on \(\lambda_{2}\). This gives the
factorisation. Each factor is the likelihood of grouped exponential
data:
\(\log(e^{-\lambda a}-e^{-\lambda b})=-\lambda a+\log(1-e^{-\lambda(b-a)})\)
is strictly concave in \(\lambda\) for \(0\leq a<b\), and the survival
term is linear, so the log-likelihood is strictly concave when some
count falls in a cell of failures; it tends to \(-\infty\) as
\(\lambda\to0\) if there is a failure, and as \(\lambda\to\infty\) if
some count lies outside the first cell. The conditions state exactly
this for both factors, and
\((\lambda_{1},\lambda_{2})\mapsto\boldsymbol{\theta}\) is a bijection.
\(\square\)

\emph{Proof of Lemma 2.} (i)
\(\boldsymbol{1}^{\top}\boldsymbol{\Sigma}(\boldsymbol{\theta}^{\ast})=\boldsymbol{\pi}^{\top}-(\boldsymbol{1}^{\top}\boldsymbol{\pi})\boldsymbol{\pi}^{\top}=\boldsymbol{0}^{\top}\),
so the range is contained in \(\mathcal{S}_{0}\);
\(\boldsymbol{\Sigma}(\boldsymbol{\theta}^{\ast})\boldsymbol{a}=\boldsymbol{0}\)
means \(\pi_{j}a_{j}=\pi_{j}\boldsymbol{\pi}^{\top}\boldsymbol{a}\) for
all \(j\), so by (A2) \(\boldsymbol{a}\) is a multiple of
\(\boldsymbol{1}\), and
\(\boldsymbol{\Sigma}(\boldsymbol{\theta}^{\ast})\boldsymbol{1}=\boldsymbol{0}\).
The rank is \(L=\dim\mathcal{S}_{0}\). (ii)
\(\boldsymbol{L}_{\beta}\boldsymbol{J}(\boldsymbol{\theta}^{\ast})=\boldsymbol{I}_{s}\)
gives
\(\boldsymbol{M}_{\beta}\boldsymbol{J}(\boldsymbol{\theta}^{\ast})=\boldsymbol{0}\)
and
\((\boldsymbol{J}(\boldsymbol{\theta}^{\ast})\boldsymbol{L}_{\beta})^{2}=\boldsymbol{J}(\boldsymbol{\theta}^{\ast})\boldsymbol{L}_{\beta}\);
\(\boldsymbol{D}^{\beta-1}(\boldsymbol{\theta}^{\ast})\boldsymbol{J}(\boldsymbol{\theta}^{\ast})\boldsymbol{L}_{\beta}\)
is symmetric, which gives the orthogonality;
\(\boldsymbol{1}^{\top}\boldsymbol{J}(\boldsymbol{\theta}^{\ast})=\boldsymbol{0}^{\top}\)
because
\(\boldsymbol{1}^{\top}\boldsymbol{\pi}(\boldsymbol{\theta})=1\). (iii)
\(\boldsymbol{\Gamma}_{\beta}=\boldsymbol{B}\boldsymbol{B}^{\top}\) with
\(\boldsymbol{B}=\boldsymbol{C}\boldsymbol{M}_{\beta}\boldsymbol{\Sigma}^{1/2}(\boldsymbol{\theta}^{\ast})\),
whose range is \(\boldsymbol{C}\boldsymbol{M}_{\beta}(\mathcal{S}_{0})\)
by (i). \(\boldsymbol{M}_{\beta}\) maps \(\mathcal{S}_{0}\) into itself
by (ii), with kernel the column space of
\(\boldsymbol{J}(\boldsymbol{\theta}^{\ast})\), which lies in
\(\mathcal{S}_{0}\) and has dimension \(s\); so
\(\dim\boldsymbol{M}_{\beta}(\mathcal{S}_{0})=L-s\). Finally
\(\boldsymbol{C}\) is injective on \(\mathcal{S}_{0}\): if
\(\boldsymbol{C}\boldsymbol{v}=\boldsymbol{0}\) the first \(L\) partial
sums of \(\boldsymbol{v}\) vanish, so \(v_{1}=\cdots=v_{L}=0\), and
\(\boldsymbol{1}^{\top}\boldsymbol{v}=0\) gives \(v_{L+1}=0\).
\(\square\)

\emph{Proof of Theorems 1 and 2.} Theorem 1 is the case
\(\boldsymbol{d}=\boldsymbol{0}\) of Theorem 2, except for the last
statement. Let
\(\boldsymbol{p}_{n}=\boldsymbol{\pi}(\boldsymbol{\theta}^{\ast})+\boldsymbol{d}/\sqrt{n}\).
\(\widehat{\boldsymbol{p}}\) is the mean of \(n\) independent indicator
vectors with mean \(\boldsymbol{p}_{n}\) and covariance
\(\boldsymbol{D}(\boldsymbol{p}_{n})-\boldsymbol{p}_{n}\boldsymbol{p}_{n}^{\top}\xrightarrow[n\to\infty]{}\boldsymbol{\Sigma}(\boldsymbol{\theta}^{\ast})\);
they are bounded, so the Lindeberg--Feller theorem gives
\(\sqrt{n}(\widehat{\boldsymbol{p}}-\boldsymbol{p}_{n})\xrightarrow[n\to\infty]{\mathcal{L}}\mathcal{N}_{L+1}(\boldsymbol{0},\boldsymbol{\Sigma}(\boldsymbol{\theta}^{\ast}))\),
hence
\(\sqrt{n}(\widehat{\boldsymbol{p}}-\boldsymbol{\pi}(\boldsymbol{\theta}^{\ast}))\xrightarrow[n\to\infty]{\mathcal{L}}\mathcal{N}_{L+1}(\boldsymbol{d},\boldsymbol{\Sigma}(\boldsymbol{\theta}^{\ast}))\),
and Proposition 2 applies. A Taylor expansion of
\(\boldsymbol{\pi}(\widehat{\boldsymbol{\theta}}_{\beta})\) around
\(\boldsymbol{\theta}^{\ast}\), with remainder
\(R(\|\widehat{\boldsymbol{\theta}}_{\beta}-\boldsymbol{\theta}^{\ast}\|)\|\widehat{\boldsymbol{\theta}}_{\beta}-\boldsymbol{\theta}^{\ast}\|\),
and Proposition 2 give \[
\boldsymbol{Y}_{n}\overset{\Delta}{=}\sqrt{n}\big(\widehat{\boldsymbol{F}}-\boldsymbol{F}_{\widehat{\boldsymbol{\theta}}_{\beta}}\big)=\boldsymbol{C}\boldsymbol{M}_{\beta}\sqrt{n}\big(\widehat{\boldsymbol{p}}-\boldsymbol{\pi}(\boldsymbol{\theta}^{\ast})\big)+\boldsymbol{R}_{n},\qquad\boldsymbol{R}_{n}\xrightarrow[n\to\infty]{P}\boldsymbol{0},
\] so that \[
\boldsymbol{Y}_{n}\xrightarrow[n\to\infty]{\mathcal{L}}\boldsymbol{Y}\sim\mathcal{N}_{L}\big(\boldsymbol{C}\boldsymbol{M}_{\beta}\boldsymbol{d},\boldsymbol{\Gamma}_{\beta}\big).
\] Since
\(\widehat{\boldsymbol{\theta}}_{\beta}\xrightarrow[n\to\infty]{P}\boldsymbol{\theta}^{\ast}\)
and \(\boldsymbol{W}(\boldsymbol{\theta})\) is continuous,
\(|\boldsymbol{Y}_{n}^{\top}(\boldsymbol{W}(\widehat{\boldsymbol{\theta}}_{\beta})-\boldsymbol{W}(\boldsymbol{\theta}^{\ast}))\boldsymbol{Y}_{n}|\leq\|\boldsymbol{Y}_{n}\|^{2}\|\boldsymbol{W}(\widehat{\boldsymbol{\theta}}_{\beta})-\boldsymbol{W}(\boldsymbol{\theta}^{\ast})\|\xrightarrow[n\to\infty]{P}0\),
so
\(T_{w,\beta}\xrightarrow[n\to\infty]{\mathcal{L}}\boldsymbol{Y}^{\top}\boldsymbol{W}(\boldsymbol{\theta}^{\ast})\boldsymbol{Y}\).
Let
\(\boldsymbol{V}=\boldsymbol{W}^{1/2}(\boldsymbol{\theta}^{\ast})\boldsymbol{Y}\),
with covariance
\(\boldsymbol{W}^{1/2}\boldsymbol{\Gamma}_{\beta}\boldsymbol{W}^{1/2}=\sum_{j\leq L-s}\lambda_{j}\boldsymbol{q}_{j}\boldsymbol{q}_{j}^{\top}\)
(Lemma 2(iii), and \(\boldsymbol{W}(\boldsymbol{\theta}^{\ast})\) is
positive definite), and mean
\(\boldsymbol{W}^{1/2}\boldsymbol{C}\boldsymbol{M}_{\beta}\boldsymbol{d}\).
Because \(\boldsymbol{d}\in\mathcal{S}_{0}\),
\(\boldsymbol{C}\boldsymbol{M}_{\beta}\boldsymbol{d}\in\boldsymbol{C}\boldsymbol{M}_{\beta}(\mathcal{S}_{0})=\operatorname{range}\boldsymbol{\Gamma}_{\beta}\),
so the mean of \(\boldsymbol{V}\) lies in the span of
\(\boldsymbol{q}_{1},\ldots,\boldsymbol{q}_{L-s}\). Completing them to
an orthonormal basis, the components of \(\boldsymbol{V}\) on the added
vectors have mean and variance zero, and
\(\boldsymbol{q}_{j}^{\top}\boldsymbol{V}=\lambda_{j}^{1/2}(Z_{j}+\delta_{j})\)
with independent \(Z_{j}\sim\mathcal{N}(0,1)\). Hence
\(\boldsymbol{Y}^{\top}\boldsymbol{W}\boldsymbol{Y}=\|\boldsymbol{V}\|^{2}=\sum_{j}\lambda_{j}(Z_{j}+\delta_{j})^{2}\).
If
\(\boldsymbol{d}=\boldsymbol{J}(\boldsymbol{\theta}^{\ast})\boldsymbol{h}\),
\(\boldsymbol{M}_{\beta}\boldsymbol{d}=\boldsymbol{0}\) by Lemma 2(ii).
The limit of Theorem 1 is chi-square if and only if all \(\lambda_{j}\)
equal one, that is,
\(\boldsymbol{W}^{1/2}\boldsymbol{\Gamma}_{\beta}\boldsymbol{W}^{1/2}\)
is idempotent, which is equivalent to
\(\boldsymbol{\Gamma}_{\beta}\boldsymbol{W}\boldsymbol{\Gamma}_{\beta}=\boldsymbol{\Gamma}_{\beta}\).
\(\square\)

\emph{Proof of the statement of Remark 3 (the other members of the
family).} In this proof the dependence on \(n\) is written explicitly.
Let \(\widehat{F}_{n,j}\) be the empirical distribution function at
\(t_{j}\),
\(\widetilde{F}_{n,j}=F_{\widehat{\boldsymbol{\theta}}_{\beta}}(t_{j})\)
the fitted one, \(r_{n,j}=\widehat{F}_{n,j}-\widetilde{F}_{n,j}\) and
\(F^{\ast}_{j}=F_{\boldsymbol{\theta}^{\ast}}(t_{j})\in(0,1)\),
\(j=1,\ldots,L\). By the proof of Theorems 1 and 2,
\(\sqrt{n}\,r_{n,j}\) is bounded in probability, under \(H_{0}\) and
under (4), and
\(\widetilde{F}_{n,j}\xrightarrow[n\to\infty]{P}F^{\ast}_{j}\) by
Proposition 2.

Consider first \(T^{\mathrm{C}}_{\phi}\) and let
\(\varepsilon_{n,j}=r_{n,j}/\widetilde{F}_{n,j}\), so that
\(\widehat{F}_{n,j}/\widetilde{F}_{n,j}=1+\varepsilon_{n,j}\),
\(\varepsilon_{n,j}\geq-1\) and
\(\varepsilon_{n,j}\xrightarrow[n\to\infty]{P}0\). Since
\(\phi^{\ast}(1)=(\phi^{\ast})'(1)=0\) and \((\phi^{\ast})''=\phi''\) is
continuous in a neighbourhood of \(1\), Taylor's theorem gives \[
\phi^{\ast}(1+x)=\frac{\phi''(1)}{2}\,x^{2}+R(x)\,x^{2},\qquad x\geq-1,
\] where \(R\) is a deterministic function, defined by
\(R(x)=\phi^{\ast}(1+x)/x^{2}-\phi''(1)/2\) for \(x\neq0\) and
\(R(0)=0\), with \(R(x)\to0\) as \(x\to0\). Substituting
\(x=\varepsilon_{n,j}\) in the definition of \(T^{\mathrm{C}}_{\phi}\),
\[
T^{\mathrm{C}}_{\phi}=\frac{2n}{\phi''(1)}\sum_{j=1}^{L}\widetilde{F}_{n,j}\,\phi^{\ast}(1+\varepsilon_{n,j})
=n\sum_{j=1}^{L}\frac{r_{n,j}^{2}}{\widetilde{F}_{n,j}}+\frac{2}{\phi''(1)}\sum_{j=1}^{L}\frac{n\,r_{n,j}^{2}}{\widetilde{F}_{n,j}}\,R(\varepsilon_{n,j}).
\] The first sum is \(T_{w,\beta}\) with \(w(F)=1/F\). In the second,
\(n\,r_{n,j}^{2}/\widetilde{F}_{n,j}\) is bounded in probability,
because \(\sqrt{n}\,r_{n,j}\) is and
\(\widetilde{F}_{n,j}\xrightarrow[n\to\infty]{P}F^{\ast}_{j}>0\), and
\(R(\varepsilon_{n,j})\xrightarrow[n\to\infty]{P}0\), because \(R\) is
continuous at \(0\) and
\(\varepsilon_{n,j}\xrightarrow[n\to\infty]{P}0\). Hence \[
T^{\mathrm{C}}_{\phi}-n\sum_{j=1}^{L}\frac{r_{n,j}^{2}}{\widetilde{F}_{n,j}}=\frac{2}{\phi''(1)}\sum_{j=1}^{L}\frac{n\,r_{n,j}^{2}}{\widetilde{F}_{n,j}}\,R(\varepsilon_{n,j})\xrightarrow[n\to\infty]{P}0 .
\] The survival member is treated in the same way with
\(1-\widehat{F}_{n,j}-(1-\widetilde{F}_{n,j})=-r_{n,j}\) and
\(1-\widetilde{F}_{n,j}\xrightarrow[n\to\infty]{P}1-F^{\ast}_{j}>0\),
and \(T^{\mathrm{CS}}_{\phi}\) is the sum of both, with
\(1/F+1/(1-F)=1/(F(1-F))\).

For \(T^{\mathrm{C}}_{a}\), let
\(g(x,y)=y^{1+a}-(1+a^{-1})y^{a}x+a^{-1}x^{1+a}\) for \(x\geq0\),
\(y>0\). Then \(g(y,y)=0\), \(\partial g(x,y)/\partial x|_{x=y}=0\) and
\(\partial^{2}g(x,y)/\partial x^{2}=(1+a)x^{a-1}\), which is continuous
on \((0,\infty)\). By Taylor's theorem, \[
g(y+h,y)=\frac{1+a}{2}\,y^{a-1}h^{2}+R_{a}(h,y)\,h^{2},
\] where \(R_{a}(h,y)=g(y+h,y)/h^{2}-\frac{1}{2}(1+a)y^{a-1}\) for
\(h\neq0\) and \(R_{a}(0,y)=0\) is a deterministic function with
\(R_{a}(h,y)\to0\) as \((h,y)\to(0,y^{\ast})\), for every
\(y^{\ast}>0\); this follows from the integral form of the remainder,
\(R_{a}(h,y)=\int_{0}^{1}(1-s)(\partial^{2}g(y+sh,y)/\partial x^{2}-(1+a)y^{a-1})\,ds\),
and the continuity of \(x\mapsto x^{a-1}\) at \(y^{\ast}\). Hence \[
T^{\mathrm{C}}_{a}-n\sum_{j=1}^{L}\widetilde{F}_{n,j}^{\,a-1}r_{n,j}^{2}=\frac{2}{1+a}\sum_{j=1}^{L}n\,r_{n,j}^{2}\,R_{a}(r_{n,j},\widetilde{F}_{n,j})\xrightarrow[n\to\infty]{P}0,
\] because \(n\,r_{n,j}^{2}\) is bounded in probability and
\((r_{n,j},\widetilde{F}_{n,j})\xrightarrow[n\to\infty]{P}(0,F^{\ast}_{j})\).
The first sum is \(T_{w,\beta}\) with \(w(F)=F^{a-1}\). \(\square\)

\emph{Proof of Corollary 1.} The nonzero eigenvalues of
\(\boldsymbol{\Gamma}_{\beta}\boldsymbol{W}(\boldsymbol{\theta}^{\ast})\)
are continuous functions of \(\boldsymbol{\theta}^{\ast}\), through
\(\boldsymbol{J}\), \(\boldsymbol{D}\) and \(\boldsymbol{W}\), and their
number \(L-s\) does not change near \(\boldsymbol{\theta}^{\ast}\); the
\(1-\alpha\) quantile of \(\sum_{j}\lambda_{j}Z_{j}^{2}\) is a
continuous function of positive weights, because the distribution is
continuous and strictly increasing on \((0,\infty)\). Hence
\(\widehat{q}_{1-\alpha}\xrightarrow[n\to\infty]{P} q_{1-\alpha}\),
under \(H_{0}\) and under the contiguous alternatives (4), by
Proposition 2. By Slutsky's theorem,
\(T_{w,\beta}-\widehat{q}_{1-\alpha}\) converges in distribution to the
limit of Theorem 1 or 2 minus \(q_{1-\alpha}\), and the limit
distributions are continuous. \(\square\)

\emph{Proof of Proposition 4.}
\(\widehat{\boldsymbol{p}}\xrightarrow[n\to\infty]{P}\boldsymbol{p}\)
and
\(\widehat{\boldsymbol{\theta}}_{\beta}\xrightarrow[n\to\infty]{P}\boldsymbol{\theta}_{\beta}(\boldsymbol{p})\),
so
\(\widehat{\boldsymbol{F}}-\boldsymbol{F}_{\widehat{\boldsymbol{\theta}}_{\beta}}\xrightarrow[n\to\infty]{P}\boldsymbol{C}(\boldsymbol{p}-\boldsymbol{\pi}(\boldsymbol{\theta}_{\beta}(\boldsymbol{p})))\),
which is not zero because
\(\boldsymbol{p}-\boldsymbol{\pi}(\boldsymbol{\theta}_{\beta}(\boldsymbol{p}))\)
is a nonzero vector of \(\mathcal{S}_{0}\) and \(\boldsymbol{C}\) is
injective on \(\mathcal{S}_{0}\). Since
\(\boldsymbol{W}(\boldsymbol{\theta}_{\beta}(\boldsymbol{p}))\) is
positive definite, \(T_{w,\beta}/n\) converges in probability to a
positive constant, while \(\widehat{q}_{1-\alpha}\) stays bounded in
probability. \(\square\)

\emph{Proof of Proposition 5.}
\(\boldsymbol{G}_{\varepsilon}\in\mathcal{P}_{L+1}\) and
\(\boldsymbol{G}_{\varepsilon}\to\boldsymbol{\pi}(\boldsymbol{\theta}^{\ast})\)
as \(\varepsilon\to0\), so
\(\boldsymbol{G}_{\varepsilon}\in\mathcal{V}\) for \(\varepsilon\)
small, and by Proposition 1 the map
\(\varepsilon\mapsto\boldsymbol{T}_{\beta}(\boldsymbol{G}_{\varepsilon})\)
is continuously differentiable, with value
\(\boldsymbol{\theta}^{\ast}\) and derivative
\(\boldsymbol{L}_{\beta}(\boldsymbol{q}-\boldsymbol{\pi}(\boldsymbol{\theta}^{\ast}))\)
at \(\varepsilon=0\). Hence
\(\boldsymbol{r}_{\varepsilon}=\boldsymbol{r}(\boldsymbol{G}_{\varepsilon})\)
satisfies \(\boldsymbol{r}_{0}=\boldsymbol{0}\) and
\(\boldsymbol{r}_{\varepsilon}=\varepsilon\boldsymbol{h}_{\beta}(\boldsymbol{q})+\boldsymbol{R}(\varepsilon)\varepsilon\),
with \(\boldsymbol{R}(\varepsilon)\to\boldsymbol{0}\) as
\(\varepsilon\to0\), because the derivative of
\(\boldsymbol{r}_{\varepsilon}\) at \(0\) is
\(\boldsymbol{C}(\boldsymbol{q}-\boldsymbol{\pi}(\boldsymbol{\theta}^{\ast})-\boldsymbol{J}(\boldsymbol{\theta}^{\ast})\boldsymbol{L}_{\beta}(\boldsymbol{q}-\boldsymbol{\pi}(\boldsymbol{\theta}^{\ast})))=\boldsymbol{h}_{\beta}(\boldsymbol{q})\).
Let
\(\boldsymbol{W}_{\varepsilon}=\boldsymbol{W}(\boldsymbol{T}_{\beta}(\boldsymbol{G}_{\varepsilon}))\),
which is continuous in \(\varepsilon\). Then \[
g(\varepsilon)=\varepsilon^{2}\big(\boldsymbol{h}_{\beta}(\boldsymbol{q})+\boldsymbol{R}(\varepsilon)\big)^{\top}\boldsymbol{W}_{\varepsilon}\big(\boldsymbol{h}_{\beta}(\boldsymbol{q})+\boldsymbol{R}(\varepsilon)\big)=\varepsilon^{2}\,\boldsymbol{h}_{\beta}^{\top}(\boldsymbol{q})\boldsymbol{W}(\boldsymbol{\theta}^{\ast})\boldsymbol{h}_{\beta}(\boldsymbol{q})+R_{1}(\varepsilon)\varepsilon^{2},
\] with \(R_{1}(\varepsilon)\to0\) as \(\varepsilon\to0\), which gives
\(g(0)=g'(0)=0\) and the value of \(g''(0)\); the derivatives of
\(\boldsymbol{W}\) and the second derivative of
\(\boldsymbol{T}_{\beta}\) do not enter, because they multiply
\(\boldsymbol{r}_{0}=\boldsymbol{0}\). For Pearson's statistic the
argument is the same without \(\boldsymbol{C}\). \(\square\)

\end{document}